\documentclass[showpacs,aps,graphicx,twocolumn]{revtex4}
\usepackage{amssymb}
\usepackage{amsmath}%preprint,
\usepackage{graphicx}
\usepackage{multirow}

\begin{document}

%\begin{CJK*}{GBK}{song} %显示中文

\title{Hyperentanglement purification for two-photon six-qubit quantum systems\footnote{Published in Phys. Rev. A \textbf{94}, 032319 (2016)}}

\author{Guan-Yu Wang, Qian Liu, and Fu-Guo Deng\footnote{Corresponding author: fgdeng@bnu.edu.cn} }

\address{Department of Physics, Applied Optics Beijing Area Major Laboratory,
Beijing Normal University, Beijing 100875, China}

\date{\today }

\begin{abstract}
Recently, two-photon six-qubit hyperentangled states were produced in
experiment and they can improve the channel capacity of quantum
communication largely. Here we present a scheme for the
hyperentanglement purification of nonlocal two-photon systems in
three degrees of freedom (DOFs), including the polarization, the
first-longitudinal-momentum, and the second longitudinal momentum
DOFs. Our hyperentanglement purification protocol (hyper-EPP) is constructed with two steps resorting to
parity-check quantum nondemolition measurement on the three DOFs and
SWAP gates, respectively. With these two steps, the bit-flip errors
in the three DOFs can be corrected efficiently. Also, we show that using SWAP gates is
a universal method for hyper-EPP in the polarization DOF and
multiple longitudinal momentum DOFs. The implementation of our
hyper-EPP is assisted by nitrogen-vacancy centers in optical
microcavities, which could be achieved with current techniques. It
is useful for long-distance high-capacity quantum communication with
two-photon six-qubit hyperentanglement.
\end{abstract}

 \maketitle

\section{Introduction}
\label{sec1}

Quantum entanglement plays a critical role in quantum information
processing \cite{quantum1}. It is the key resource in quantum
communication, such as   quantum teleportation \cite{tele1}, quantum
dense coding \cite{dense1,super2}, quantum key distribution
\cite{Ekert,BBM92}, quantum secret sharing \cite{QSS1}, quantum
secure direct communication \cite{QSDC1,twostep}, and so on.
Maximally entangled states can be used as the quantum channel for
teleporting an unknown state of a quantum particle, without moving
the particle itself  \cite{tele1}. It can also be used to carry two
bits of information by moving only one two-level particle
\cite{dense1}, not two or more. The two parties of quantum
communication can create a private key with entangled photon pairs
in an absolutely secure way \cite{Ekert,BBM92}. Moreover, they can
exchange the secret message directly and securely without
distributing the private key if they share maximally entangled
photon pairs \cite{QSDC1,twostep}.

Hyperentanglement, a state of a quantum system being simultaneously
entangled in multiple degrees of freedom (DOFs),  has attracted much
attention as it can improve both the channel capacity and the
security  of quantum communication largely, beat the channel
capacity of linear photonic superdense coding \cite{super1},  assist
the complete analysis of Bell states \cite{BSA1,BSA2,BSA3,BSA4}, and
so on.  With the $\beta$ barium borate (BBO) crystal, photon pairs
produced by spontaneous parametric down conversion (PDC) can be in a
hyperentangled state. Many theoretical and experimental schemes  for
the generation of hyperentangled states have been proposed and
implemented in optical systems
\cite{hyper1,hyper2,hyper3,hyper4,hyper5,hyper6,hyper7,hyper8,hyper9},
such as in polarization-momentum DOFs \cite{hyper2},
polarization-orbital-angular momentum DOFs \cite{hyper4}, multipath
DOFs \cite{hyper5},  and so on. In 2009,
Vallone \emph{et al.} \cite{hyper6} demonstrated experimentally the
generation of a two-photon six-qubit hyperentangled state in three
DOFs.

%polarization-frequency DOFs \cite{hyper7},
%polarization-time-bin DOFs \cite{hyper9},

Although the interaction between a photon and its environment is
weaker than other quantum systems, it still inevitably suffers from
channel noise, such as thermal fluctuation, vibration, imperfection
of an optical fiber, and birefringence effects. The interaction
between the photons and the environment will make the entangled
photon pairs in less entangled states or even in mixed states,
which will decrease the security and the efficiency of quantum
communication. Entanglement purification is used to obtain a subset
of high-fidelity nonlocal entangled quantum systems from a set of
those in mixed entangled states \cite{EPP1,EPP2,EPP3,EPP4,EPP5}. In
1996, Bennett \emph{et al.} \cite{EPP1} introduced the  entanglement purification protocol (EPP) for quantum systems in a
Werner state \cite{Werner} with quantum controlled-NOT   gates. In
2001, Pan \emph{et al.} \cite{EPP3} proposed an EPP  with linear
optical elements. In 2002, Simon and Pan \cite{EPP4} presented an
EPP for a PDC source, not an ideal entanglement source. In 2008,
Sheng \emph{et al.} \cite{EPP5} proposed an efficient EPP for
polarization entanglement from a PDC source, assisted by
nondestructive quantum nondemolition detectors (QND) with cross-Kerr
nonlinearity. In 2010, Sheng and Deng \cite{EPP6} introduced the
original deterministic EPP for two-photon systems, which is far
different from the conventional EPPs \cite{EPP1,EPP2,EPP3,EPP4,EPP5}
as it works in a deterministic way, not a probabilistic one.
Subsequently, some interesting deterministic EPPs were proposed
\cite{EPP7,EPP8,EPP9,EPP10}. In 2003, Pan \emph{et al.} \cite{EPP11}
demonstrated the  entanglement purification of photon pairs using
linear optical elements.   Recently, Ren and Deng \cite{EPP16}
proposed the original hyperentanglement purification protocol
(hyper-EPP) for two-photon four-qubit systems in mixed
polarization-spatial hyperentangled Bell states with polarization
bit-flip errors and spatial-mode bit-flip errors assisted by
nonlinear optical elements.

A nitrogen vacancy (NV) center in diamond is an attractive candidate
for quantum information processing because of its long-lived
coherence time at room temperature and optical controllability. The
coherence time of a diamond NV center can continue for 1.8 ms
\cite{time}. In a diamond NV center, the electron spin can be
exactly populated by the optical pumping with 532-nm light, and it
can be easily manipulated \cite{m1,m2,m3,m4} and readout
\cite{r1,r2} by using the microwave excitation. Many interesting
approaches based on an NV center in diamond coupled to an optical
cavity have been proposed in theory \cite{NV1,NV2,NV3,NV4,NV7} and
implemented in experiment \cite{NV8,NV9,NV10,NV11,NV12}.

In this paper, we present a hyper-EPP for nonlocal photon
systems entangled in three DOFs assisted by nitrogen-vacancy centers
in optical microcavities, including the polarization DOF, the first-longitudinal-momentum DOF, and the second-longitudinal-momentum DOF.
Our protocol is completed by two steps. The first step resorts to
parity-check quantum nondemolition measurement on the polarization
DOF (P-QND) and the two longitudinal-momentum DOFs (S-QND). The
second step resorts to the SWAP gate between the polarization states
of two photons (P-P-SWAP gate) and the SWAP gate between the
polarization state and the spatial state of one photon (P-F-SWAP
gate and P-S-SWAP gate). Also, we show that using the SWAP gates is a universal method
for hyper-EPP in the polarization DOF and multiple longitudinal-momentum DOFs. Our hyper-EPP can effectively improve the
entanglement of photon systems in long-distance quantum
communication.

This paper is organized as follows: We give the interface between a
circularly polarized light and  a diamond nitrogen-vacancy center
confined in an optical microcavity in Sec. \ref{sec2}. Subsequently,
we present the polarization-spatial parity-check QND of two-photon
six-qubit systems in Sec. \ref{sec3}, and then, we give the
principle of our SWAP gate between the polarization states of two
photons in Sec. \ref{sec41} and that of our SWAP gate between the
polarization state and the spatial state of one photon in Sec.
\ref{sec42}. In Sec.\ref{sec5}, we propose an efficient hyper-EPP
for mixed two-photon six-qubit hyperentangled Bell states. In Sec.
\ref{sec6}, we discuss the expansion for purifying the two-photon
systems in the polarization DOF and multiple longitudinal momentum
DOFs. A discussion and a summary are given in Sec. \ref{sec7}. In
addition, some other cases for our hyper-EPP are discussed in the
Appendix.

\section{The interface between a circularly polarized light and  a diamond nitrogen-vacancy center confined in
an optical microcavity} \label{sec2}

\begin{figure}[th]%[tpb]
\centering
\includegraphics[width=8cm,angle=0]{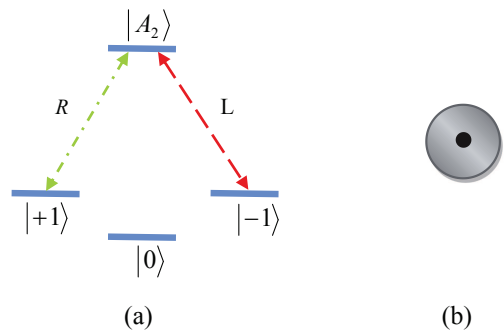}
\caption{(a) The energy-level diagram of an NV-cavity system.
The qubit is encoded on the ground states $|\pm1\rangle$ and the excited state $|A_{2}\rangle$.
The transitions $|\pm1\rangle\leftrightarrow|A_{2}\rangle$ are driven by the right and light circularly polarized photons,
respectively. (b) Schematic diagram for an NV center confined inside a single-sided optical resonant microcavity.} \label{fig1}
\end{figure}

A diamond NV center consists of a vacancy adjacent to a
substitutional nitrogen atom. The NV center is negatively charged
with two unpaired electrons located at the vacancy. The energy-level
structure of the NV center coupled to the cavity mode is shown in
Fig.~\ref{fig1}(a). The ground state is a spin triplet with the
splitting at 2.87 GHz in a zero external field between levels
$|0\rangle\equiv|m=0\rangle$ and $|\pm1\rangle\equiv|m=\pm1\rangle$
owing to spin-spin interaction. The excited state, which is one of
the six eigenstates of the full Hamiltonian including spin-spin and
spin-orbit interactions in the absence of any perturbation, is
labeled as
$|A_{2}\rangle=|E_{-}\rangle|+1\rangle+|E_{+}\rangle|-1\rangle$,
where $|E_{+}\rangle$ and $|E_{-}\rangle$ are the orbital states
with the angular momentum projections $+1$ and $-1$ along the NV
axis, respectively. We encode the qubits on the sublevels
$|\pm1\rangle$ as the ground states, and take $|A_{2}\rangle$ as an
auxiliary excited state. The transitions $|+1\rangle
\leftrightarrow |A_{2}\rangle$ and $|-1\rangle
\leftrightarrow |A_{2}\rangle$ in the NV center are resonantly
coupled to the right (R) and the left (L) circularly polarized
photons with the identical transition frequency, respectively. The
two transitions take place with   equal probability.

Let us consider the composite unit, a diamond NV center confined
inside a single-sided resonant microcavity, as shown in
Fig.~\ref{fig1}(b). The Heisenberg equations of motion for this
system can be written as \cite{Walls}
\begin{eqnarray}
\begin{split}
\dot{a}(t)\;=\; &-\left[i(\omega_{c}-\omega_{p})+\frac{\kappa}{2}\right]a(t)-g\sigma_{-}(t)\\
&-\sqrt{\kappa}\,a_{in},\\
\dot{\sigma}_{-}(t)\;=\; &-\left[i(\omega_{0}-\omega_{p})+\frac{\gamma}{2}\right]\sigma_{-}(t)-g\sigma_{z}(t)\,a(t)\;\;\;\;\\
 &+\sqrt{\gamma}\,\sigma_{z}(t)b_{in}(t),\\
a_{out}(t)\;=\; & a_{in}(t)+\sqrt{\kappa}\,a(t).
\end{split}
\end{eqnarray}
Here $a(t)$ and $\sigma_{-}(t)$ are the annihilation operator of the
cavity mode and the transition operator of the diamond NV center.
$\sigma_{z}(t)$ represents the inversion operator of the NV center.
$\omega_{c}$, $\omega_{p}$, and $\omega_{0}$ are the frequencies of
the cavity mode, the photon, and the diamond NV center level
transition, respectively. $g$, $\gamma$, and $\kappa$ are the
coupling strength between a diamond NV center and a cavity, the
decay rate of a diamond NV center, and the damping rate of a cavity,
respectively. $b_{in}(t)$ is the vacuum input field with the
commutation relation
$[b_{in}(t),b_{in}^{\dagger}(t')]=\delta(t-t')$.

In the weak excitation limit, $\langle \sigma_{z}\rangle=-1$, the
reflection coefficient for the NV-cavity unit is \cite{An,Hu}
\begin{equation}\begin{split}
r(\omega_{p})=\frac{[i(\omega_{c}-\omega_{p})-\frac{\kappa}{2}][i(\omega_{0}-\omega_{p})
+\frac{\gamma}{2}]+g^{2}}{[i(\omega_{c}-\omega_{p})+\frac{\kappa}{2}][i(\omega_{0}-\omega_{p})+\frac{\gamma}{2}]+g^{2}}.
\end{split}\end{equation}
When the diamond NV center is uncoupled to the cavity or coupled to
an empty cavity, that is, $g=0$, the reflection coefficient can be
written as
\begin{equation}\begin{split}
r_{0}(\omega_{p})=\frac{i(\omega_{c}-\omega_{p})-\frac{\kappa}{2}}{i(\omega_{c}-\omega_{p})+\frac{\kappa}{2}}.
\end{split}\end{equation}
If $\omega_{0}=\omega_{c}=\omega_{p}$, the reflection coefficients
can be written as
\begin{equation}\begin{split}
r=\frac{-\kappa\gamma+4g^{2}}{\kappa\gamma+4g^{2}},\;\;\;\;\;\;\;\;\;\;\;r_{0}=-1.
\end{split}\end{equation}
That is, when $\omega_{0}=\omega_{c}=\omega_{p}$, the change of the input photon can be summarized as
\begin{equation}\begin{split}
&|R\rangle|+1\rangle\rightarrow r|R\rangle|+1\rangle,\;\;\;|R\rangle|-1\rangle\rightarrow -|R\rangle|-1\rangle,\\
&|L\rangle|+1\rangle\rightarrow
-|L\rangle|+1\rangle,\;\;\;|L\rangle|-1\rangle\rightarrow
r|L\rangle|-1\rangle.\;\;\;\;
\end{split}\end{equation}
Considering the condition $g\geq5\sqrt{\kappa\gamma}$, we can obtain
the approximate evolution of the system composed of a diamond NV
center and a photon as follows \cite{NV2}:
\begin{equation}\begin{split}
&|R\rangle|+1\rangle\rightarrow |R\rangle|+1\rangle,\;\;\;|R\rangle|-1\rangle\rightarrow -|R\rangle|-1\rangle,\\
&|L\rangle|+1\rangle\rightarrow
-|L\rangle|+1\rangle,\;\;\;|L\rangle|-1\rangle\rightarrow
|L\rangle|-1\rangle.\;\;\;\;
\end{split}\end{equation}

\begin{figure}[th]%[tpb]
\centering
\includegraphics[width=6cm,angle=0]{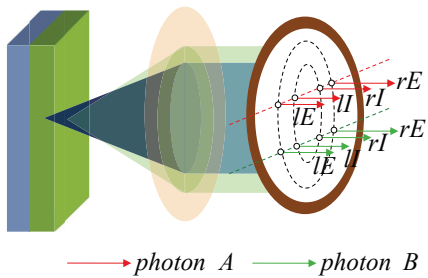}
\caption{The source of two-photon six-qubit hyperentanglement. The
detailed description is shown in \cite{hyper6}. } \label{fig2}
\end{figure}

\section{Polarization-spatial parity-check QND of two-photon six-qubit systems}
\label{sec3}

A  two-photon six-qubit hyperentangled Bell state  can be described
as follows:
\begin{eqnarray}\label{eq1}                    %  Eq 14
\begin{split}
 |\Psi\rangle_{AB}=&\frac{1}{\sqrt{2}}(|H\rangle_{A}|H\rangle_{B}+|V\rangle_{A}|V\rangle_{B})\\
 &\otimes\frac{1}{\sqrt{2}} (|l\rangle_{A}|r\rangle_{B}+|r\rangle_{A}|l\rangle_{B}) \\
 &\otimes\frac{1}{\sqrt{2}} (|I\rangle_{A}|I\rangle_{B}+|E\rangle_{A}|E\rangle_{B}).
 \end{split}
\end{eqnarray}
Here the subscripts A and B represent the two photons. $H$ and $V$
represent the horizontal and the vertical polarizations of photons,
respectively. $r$ and $l$ represent the right and the left spatial
modes of a photon in the first-longitudinal-momentum DOF,
respectively.  $E$ and $I$ denote the external and the internal
spatial modes of a photon in the second-longitudinal-momentum DOF,
respectively,  shown in Fig.~\ref{fig2}. This two-photon six-qubit
hyperentangled Bell state can be produced by  two 0.5-mm-thick
type I BBO crystal slabs aligning one behind the other and an eight-hole
screen \cite{hyper6}, shown in Fig.~\ref{fig2}.  When a
continuous-wave (cw) vertically polarized $Ar^{+}$ laser beam
interacts through spontaneous parametric downconversion (SPDC)
with the two BBO crystal slabs,  the nonlinear interaction between
the laser beam and the BBO crystal leads to the production of the
degenerate photon pairs which are entangled in polarization and
belong to the surfaces of two emission cones. As shown  in
Fig.~\ref{fig2}, the insertion of an eight-hole screen allows us to
achieve the double longitudinal-momentum entanglement.

%-$\uppercase\expandafter{\romannumeral 1}$ $\beta$ barium borate
%(

Generally speaking, an arbitrary two-photon six-qubit hyperentangled
Bell state for a photon pair $AB$ can be written as
\begin{eqnarray}\label{eq1}                    %  Eq 14
\begin{split}
 |\Psi_6\rangle_{AB}=|\psi\rangle_P\otimes|\psi\rangle_F\otimes|\psi\rangle_S,
 \end{split}
\end{eqnarray}
where $|\psi\rangle_P$ is one of the four Bell states for a
two-photon system in the polarization DOF,
\begin{equation}\begin{split}
&|\phi^{\pm}\rangle^{p}=\frac{1}{\sqrt{2}}(|RR\rangle\pm|LL\rangle)_{AC}, \\
&|\psi^{\pm}\rangle^{p}=\frac{1}{\sqrt{2}}(|RL\rangle\pm|LR\rangle)_{AC}.
\end{split}\end{equation}
$|\psi\rangle_F$ is one of the four Bell states for a two-photon
system in the first-longitudinal-momentum DOF,
\begin{equation}\begin{split}
&|\phi^{\pm}\rangle^{F}=\frac{1}{\sqrt{2}}(|rr\rangle\pm|ll\rangle)_{AC},  \\
&|\psi^{\pm}\rangle^{F}=\frac{1}{\sqrt{2}}(|rl\rangle\pm|lr\rangle)_{AC},
\end{split}\end{equation}
and $|\psi\rangle_S$ is one of the four Bell states for a two-photon
system in the second-longitudinal-momentum DOF,
\begin{equation}\begin{split}
&|\phi^{\pm}\rangle^{S}=\frac{1}{\sqrt{2}}(|EE\rangle\pm|II\rangle)_{AC},  \\
&|\psi^{\pm}\rangle^{S}=\frac{1}{\sqrt{2}}(|EI\rangle\pm|IE\rangle)_{AC}.
\end{split}\end{equation}

The polarization-spatial parity-check QND is used to distinguish the
odd-parity mode ($|\psi^{\pm}\rangle_{p}, |\psi^{\pm}\rangle_{F}$,
and $|\psi^{\pm}\rangle_{S}$) from the even-parity mode
($|\phi^{\pm}\rangle_{p}, |\phi^{\pm}\rangle_{F}$, and
$|\phi^{\pm}\rangle_{S}$) of the hyperentangled Bell states in both
the polarization and the two longitudinal-momentum DOFs.

\begin{figure}[th]%[tpb]
\centering
\includegraphics[width=8cm,angle=0]{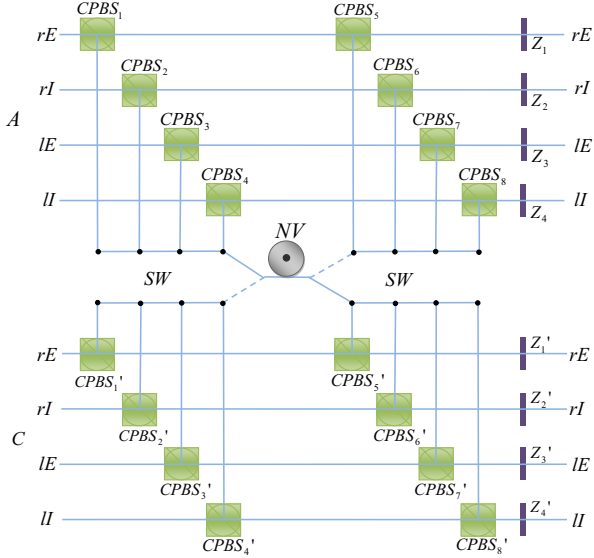}
\caption{Schematic diagram of the polarization parity-check QND.
$CPBS_{i}$ $(CPBS_{i'})$ $(i=1,2,3,4)$ is a
 circularly polarizing beam splitter
which transmits the photon in the right-circular polarization $|R\rangle$ and
reflects the photon in the left-circular polarization $|L\rangle$, respectively.
$Z_{i}$ $(Z_{i'})$ $(i=1,2,3,4)$ is a half-wave plate which performs a
polarization phase-flip operation $\sigma_{z}^{P}=|R\rangle \langle R|-|L\rangle\langle L|$ on the photon.
$SW$ is an optical switch which lets the wave packets of a photon in different spatial modes pass into and out of the NV center sequentially.} \label{fig3}
\end{figure}

The schematic diagram of the polarization parity-check QND is shown
in Fig.~\ref{fig3}, which consists of some circularly polarizing
beam splitters CPBS$_{i}$, an NV center, and some half-wave plates
Z$_{i}$. The NV center is prepared in the initial state
$|\varphi^{+}\rangle=\frac{1}{\sqrt{2}}(|+1\rangle+|-1\rangle)$. If
the two-photon system $AC$ is in a hyperentangled Bell state, one can
inject the photons $A$ and   $C$ into the quantum circuit sequentially.
The evolutions of the system consisting of the two photons and the
NV center are described as
\begin{equation}\begin{split}
&|\phi^{\pm}\rangle^{p}  |\psi\rangle_F |\psi\rangle_S \otimes |\varphi^{+}\rangle
\;\;\rightarrow\;\;
|\phi^{\pm}\rangle^{p}  |\psi\rangle_F |\psi\rangle_S \otimes |\varphi^{+}\rangle,\\
&|\psi^{\pm}\rangle^{p}  |\psi\rangle_F |\psi\rangle_S \otimes |\varphi^{+}\rangle
\;\;\rightarrow\;\;
|\psi^{\pm}\rangle^{p}  |\psi\rangle_F |\psi\rangle_S \otimes |\varphi^{-}\rangle,
\end{split}\end{equation}
where
$|\varphi^{-}\rangle=\frac{1}{\sqrt{2}}(|+1\rangle-|-1\rangle)$. By
measuring the state of the NV center in the orthogonal basis
$\{|\varphi^{+}\rangle,|\varphi^{-}\rangle\}$,  one  can distinguish
the even-parity Bell states $|\phi^{\pm}\rangle^{p}$ from the
odd-parity Bell states $|\psi^{\pm}\rangle^{p}$ of the two-photon
system in the polarization DOF without affecting the states of the
two-photon systems in the spatial-mode DOFs. That is, if the NV
center is projected into the state $|\varphi^{+}\rangle$, the
polarization state of the hyperentangled two-photon system is
$|\phi^{\pm}\rangle^{p}$. Otherwise, the polarization state of the
hyperentangled two-photon system is $|\psi^{\pm}\rangle^{p}$.

\begin{figure}[th]%[tpb]
\centering
\includegraphics[width=8cm,angle=0]{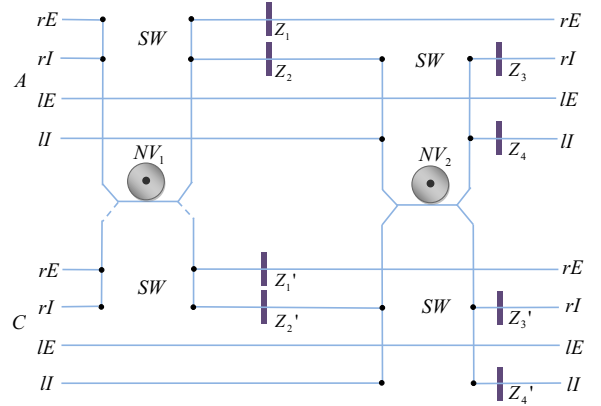}
\caption{Schematic diagram of the spatial-mode parity-check QND.} \label{fig4}
\end{figure}

The schematic diagram of the spatial-mode parity-check QND is shown
in Fig.~\ref{fig4}. It consists of two NV centers and some
half-wave plates $Z_{i}$ $(Z_{i'})$. Each of the two NV centers is
prepared in the initial state
$|\varphi^{+}\rangle_{i}=\frac{1}{\sqrt{2}}(|+1\rangle+|-1\rangle)$
with $i=1,2$. One can inject the photons $A$ and $C$ into the
quantum circuit sequentially. The evolutions of the system composed of
the two photons and the two NV centers are described as:
\begin{equation}\begin{split}
&|\psi\rangle\!_P |\phi^{\pm}\rangle\!^{F} |\phi^{\pm}\rangle\!^{S}
\!\otimes\! |\varphi\!^{+}\!\rangle_{1} |\varphi\!^{+}\!\rangle_{2}
\rightarrow
|\psi\rangle\!_P |\phi^{\pm}\rangle\!^{F} |\phi^{\pm}\rangle\!^{S} \!\otimes\! |\varphi\!^{+}\!\rangle_{1} |\varphi\!^{+}\!\rangle_{2},\\
&|\psi\rangle\!_P |\phi^{\pm}\rangle\!^{F} |\psi^{\pm}\rangle\!^{S}
\!\otimes\! |\varphi\!^{+}\!\rangle_{1} |\varphi\!^{+}\!\rangle_{2}
\rightarrow
|\psi\rangle\!_P |\phi^{\pm}\rangle\!^{F} |\psi^{\pm}\rangle\!^{S} \!\otimes\! |\varphi\!^{+}\!\rangle_{1} |\varphi\!^{-}\!\rangle_{2},\\
&|\psi\rangle\!_P |\psi^{\pm}\rangle\!^{F} |\phi^{\pm}\rangle\!^{S}
\!\otimes\! |\varphi\!^{+}\!\rangle_{1} |\varphi\!^{+}\!\rangle_{2}
\rightarrow
|\psi\rangle\!_P |\psi^{\pm}\rangle\!^{F} |\phi^{\pm}\rangle\!^{S} \!\otimes\! |\varphi\!^{-}\!\rangle_{1} |\varphi\!^{+}\!\rangle_{2},\\
&|\psi\rangle\!_P |\psi^{\pm}\rangle\!^{F} |\psi^{\pm}\rangle\!^{S}
\!\otimes\! |\varphi\!^{+}\!\rangle_{1} |\varphi\!^{+}\!\rangle_{2}
\rightarrow |\psi\rangle\!_P |\psi^{\pm}\rangle\!^{F}
|\psi^{\pm}\rangle\!^{S} \!\otimes\! |\varphi\!^{-}\!\rangle_{1}
|\varphi\!^{-}\!\rangle_{2}.
\end{split}\end{equation}

By measuring the states of two NV centers in the orthogonal basis
$\{|\varphi^{+}\rangle,|\varphi^{-}\rangle\}$, one  can distinguish
the even-parity states $|\phi^{\pm}\rangle^{F}$ from the odd-parity
states $|\psi^{\pm}\rangle^{F}$ in the first-longitudinal-momentum
DOF and the even-parity states $|\phi^{\pm}\rangle^{S}$ from the
odd-parity states $|\psi^{\pm}\rangle^{S}$ in the second
longitudinal momentum DOF, without affecting the states of the
two-photon system in the polarization DOF. That is, if $NV_{1}$ is
projected into the state $|\varphi^{+}\rangle_{1}$, the two-photon
system is in the state $|\phi^{\pm}\rangle^{F}$ in the first-longitudinal-momentum DOF and if $NV_{2}$ is projected into the
state $|\varphi^{+}\rangle_{2}$, the two-photon system is in the
state $|\phi^{\pm}\rangle^{S}$ in the second-longitudinal-momentum
DOF, respectively. If $NV_{1}$ is projected into the state
$|\varphi^{-}\rangle_{1}$, the two-photon system is in the state
$|\psi^{\pm}\rangle^{F}$ in the first-longitudinal-momentum DOF and
if $NV_{2}$ is projected into the state  $|\varphi^{-}\rangle_{2}$,
the two-photon system is in the state $|\psi^{\pm}\rangle^{S}$ in
the second-longitudinal-momentum DOF, respectively.

\begin{figure}[th]%[tpb]
\centering
\includegraphics[width=8cm,angle=0]{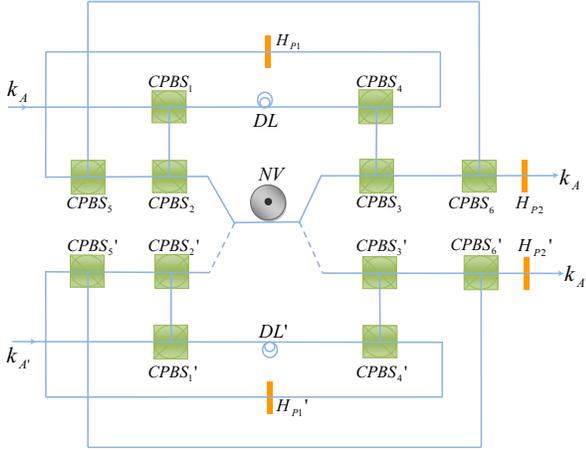}
\caption{Schematic diagram of the SWAP gate between the polarization
states of photon $A$ and photon $A'$ (P-P SWAP gate). $H_{pi}$
$(i=1,2,1',2')$ is a half-wave plate which performs the Hadamard
operation $|R\rangle \rightarrow
\frac{1}{\sqrt{2}}(|R\rangle+|L\rangle), |L\rangle \rightarrow
\frac{1}{\sqrt{2}}(|R\rangle-|L\rangle)$ on the polarization DOF of
a photon. $DL$ $(DL')$ is a time-delay device which makes the two
wave packets in different paths interfere with each other at last.
$k$ $(k=rE,\;rI,\; lE,\; lI)$ represents any spatial modes of one
photon.} \label{fig5}
\end{figure}

\section{SWAP gates}
\label{sec4}

\subsection {SWAP gate between the polarization states of two photons}
\label{sec41}

The SWAP gate between the polarization states of two photons
(P-P-SWAP gate) is used to swap the polarization states of photon
$A$ and photon $A'$ without affecting their spatial-mode states. The
initial states of two photons in the polarization DOF and two
longitudinal-momentum DOFs are
\begin{equation}\label{10}
\begin{split}
&|\Phi\rangle_{A}=|\Phi_{p}\rangle_{A} \otimes |\Phi_{k}\rangle_{A}, \\
&|\Phi\rangle_{A'}=|\Phi_{p}\rangle_{A'} \otimes |\Phi_{k}\rangle_{A'}.
\end{split}\end{equation}
Here $|\Phi_{p}\rangle_{A}=\alpha_{1}|R\rangle+\beta_{1}|L\rangle$
and $|\Phi_{p}\rangle_{A'}=\alpha_{2}|R\rangle+\beta_{2}|L\rangle$
are the polarization states of the photons $A$ and $A'$,
respectively.
$|\Phi_{k}\rangle_{A}=(\gamma_{1}|r\rangle+\delta_{1}|l\rangle)
\otimes (\varepsilon_{1}|E\rangle+\xi_{1}|I\rangle)$ and
$|\Phi_{k}\rangle_{A'}=(\gamma_{2}|r\rangle+\delta_{2}|l\rangle)
\otimes (\varepsilon_{2}|E\rangle+\xi_{2}|I\rangle)$ represent the
states of the photons $A$ and $A'$ in both the first-longitudinal-momentum DOF and the second-longitudinal-momentum DOF, respectively.
The schematic diagram of the P-P-SWAP gate is shown in
Fig.~\ref{fig5}. Suppose that the NV center is prepared in the
initial state
$|\Phi\rangle_{NV}=\frac{1}{\sqrt{2}}(|+1\rangle+|-1\rangle)$, the
SWAP gate works with the following steps.

First, one injects the photons $A$ and $A'$ into the quantum circuit
sequentially, and lets photon $A$ $(A')$ pass through the circularly
polarizing beam splitter $CPBS_{1}$ $(CPBS_{1'})$, $CPBS_{2}$
$(CPBS_{2'})$, NV center, $CPBS_{3}$ $(CPBS_{3'})$, and $CPBS_{4}$
$(CPBS_{4'})$  in sequence. After performing a Hadamard operation on
the NV center [$|+1\rangle \rightarrow
\frac{1}{\sqrt{2}}(|+1\rangle+|-1\rangle), \;|-1\rangle \rightarrow
\frac{1}{\sqrt{2}}(|+1\rangle-|-1\rangle) $], the whole state of the
system composed of two photons and one NV center is transformed
from $|\Phi\rangle_{0}$ to $|\Phi\rangle_{1}$. Here
\begin{equation}\begin{split}
|\Phi\rangle_{0}=&|\Phi\rangle_{A} \otimes |\Phi\rangle_{A'} \otimes |\Phi\rangle_{NV}, \\
|\Phi\rangle_{1}=&[\alpha_{1}\alpha_{2}|R\rangle_{A}|R\rangle_{A'}|+1\rangle   -\beta_{1}\alpha_{2}|L\rangle_{A}|R\rangle_{A'}|-1\rangle \\
&-\alpha_{1}\beta_{2}|R\rangle_{A}|L\rangle_{A'}|-1\rangle
 +\beta_{1}\beta_{2}|L\rangle_{A}|L\rangle_{A'}|+1\rangle] \\
 &\otimes |\Phi_{k}\rangle_{A} \otimes |\Phi_{k}\rangle_{A'}.
\end{split}\end{equation}

Second, one performs Hadamard operations on photons $A$ and $A'$
with the half-wave plates $H_{p1}$ and $H_{p1'}$. The state of the
whole system is changed into
\begin{equation}\begin{split}
|\Phi\rangle_{2}\,=\,&\frac{1}{2}[\alpha_{1}\alpha_{2}(|R\rangle+|L\rangle)_{A}(|R\rangle+|L\rangle)_{A'}|+1\rangle\\
&-\beta_{1}\alpha_{2}(|R\rangle-|L\rangle)_{A}(|R\rangle+|L\rangle)_{A'}|-1\rangle \\
&-\alpha_{1}\beta_{2}(|R\rangle+|L\rangle)_{A}(|R\rangle-|L\rangle)_{A'}|-1\rangle\\
&+\beta_{1}\beta_{2}(|R\rangle-|L\rangle)_{A}(|R\rangle-|L\rangle)_{A'}|+1\rangle] \\
&\otimes |\Phi_{k}\rangle_{A} \otimes |\Phi_{k}\rangle_{A'}.
\end{split}\end{equation}

Third, one lets   photon $A$ $(A')$ pass though $CPBS_{5}$
$(CPBS_{5'})$, $CPBS_{2}$ $(CPBS_{2'})$, NV center, $CPBS_{3}$
$(CPBS_{3'})$, $CPBS_{6}$ $(CPBS_{6'})$, and $H_{p2}$ $(H_{p2'})$,
and the state of the whole system is changed from $|\Phi\rangle_{2}$
to $|\Phi\rangle_{3}$. Here
\begin{equation}\begin{split}
|\Phi\rangle_{3} =&
[\alpha_{1}\alpha_{2}|R\rangle_{A}|R\rangle_{A'}|+1\rangle
-\beta_{1}\alpha_{2}|R\rangle_{A}|L\rangle_{A'}|-1\rangle \\
&-\alpha_{1}\beta_{2}|L\rangle_{A}|R\rangle_{A'}|-1\rangle
+\beta_{1}\beta_{2}|L\rangle_{A}|L\rangle_{A'}|+1\rangle] \\
&\otimes |\Phi_{k}\rangle_{A} \otimes |\Phi_{k}\rangle_{A'}.
\end{split}\end{equation}

Finally, by performing a Hadamard operation on the NV center, the
state of the whole system is transformed into
\begin{equation}\begin{split}
|\Phi\rangle_{4}=&\frac{1}{\sqrt{2}}(\alpha_{2}|R\rangle-\beta_{2}|L\rangle)_{A}|\Phi_{k}\rangle_{A}\\
& \otimes (\alpha_{1}|R\rangle-\beta_{1}|L\rangle)_{A'}|\Phi_{k}\rangle_{A'}\otimes |+1\rangle \\
&+\frac{1}{\sqrt{2}}(\alpha_{2}|R\rangle+\beta_{2}|L\rangle)_{A}|\Phi_{k}\rangle_{A} \\
& \otimes
(\alpha_{1}|R\rangle+\beta_{1}|L\rangle)_{A'}|\Phi_{k}\rangle_{A'}\otimes
|-1\rangle.
\end{split}\end{equation}
By measuring the NV center in the orthogonal basis $\{|+1\rangle,
|-1\rangle\}$, the polarization state of photon $A$ is swaped with
the polarization state of photon $A'$ without disturbing their
states in the spatial-mode DOFs. If the NV center is projected
into state $|+1\rangle$, phase-flip operations
$\sigma_{z}^{P}=|R\rangle\langle R|-|L\rangle \langle L|$ are
performed on the polarization DOF of photons $A$ and $A'$. After the
phase-flip operations, the state of the two photons is transformed
into
\begin{equation}\begin{split}
|\Phi\rangle_{f1}=&(\alpha_{2}|R\rangle+\beta_{2}|L\rangle)_{A}|\Phi_{k}\rangle_{A} \\
&\otimes
(\alpha_{1}|R\rangle+\beta_{1}|L\rangle)_{A'}|\Phi_{k}\rangle_{A'}.
\end{split}\end{equation}
Here, $|\Phi\rangle_{f1}$ is the objective state of the P-P-SWAP
gate. If the NV center is projected into state $|-1\rangle$, the
objective state can be obtained directly without phase-flip
operations.

\begin{figure}[th]%[tpb]
\centering
\includegraphics[width=8cm,angle=0]{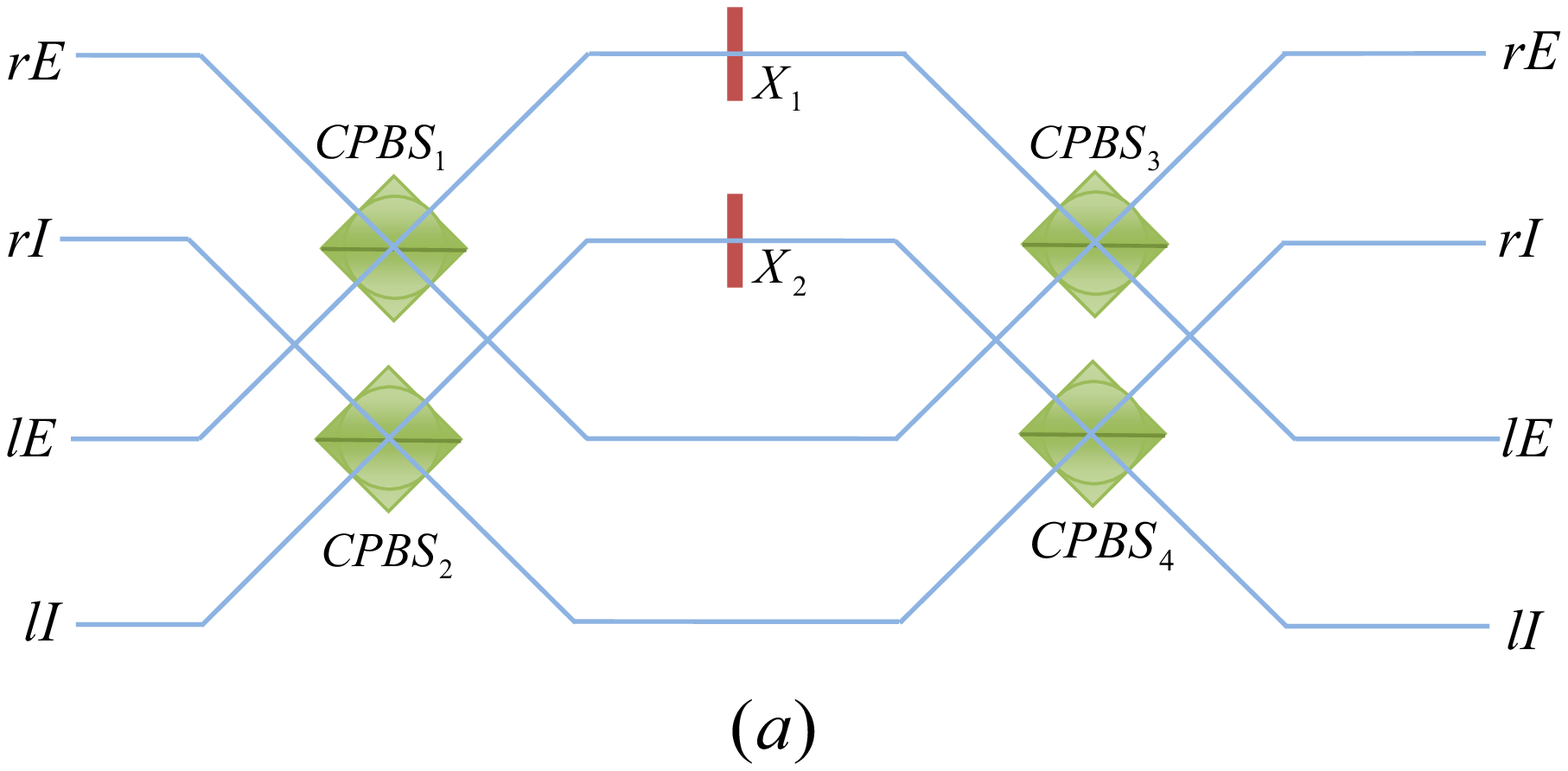} \hspace{30pt}
\includegraphics[width=8cm,angle=0]{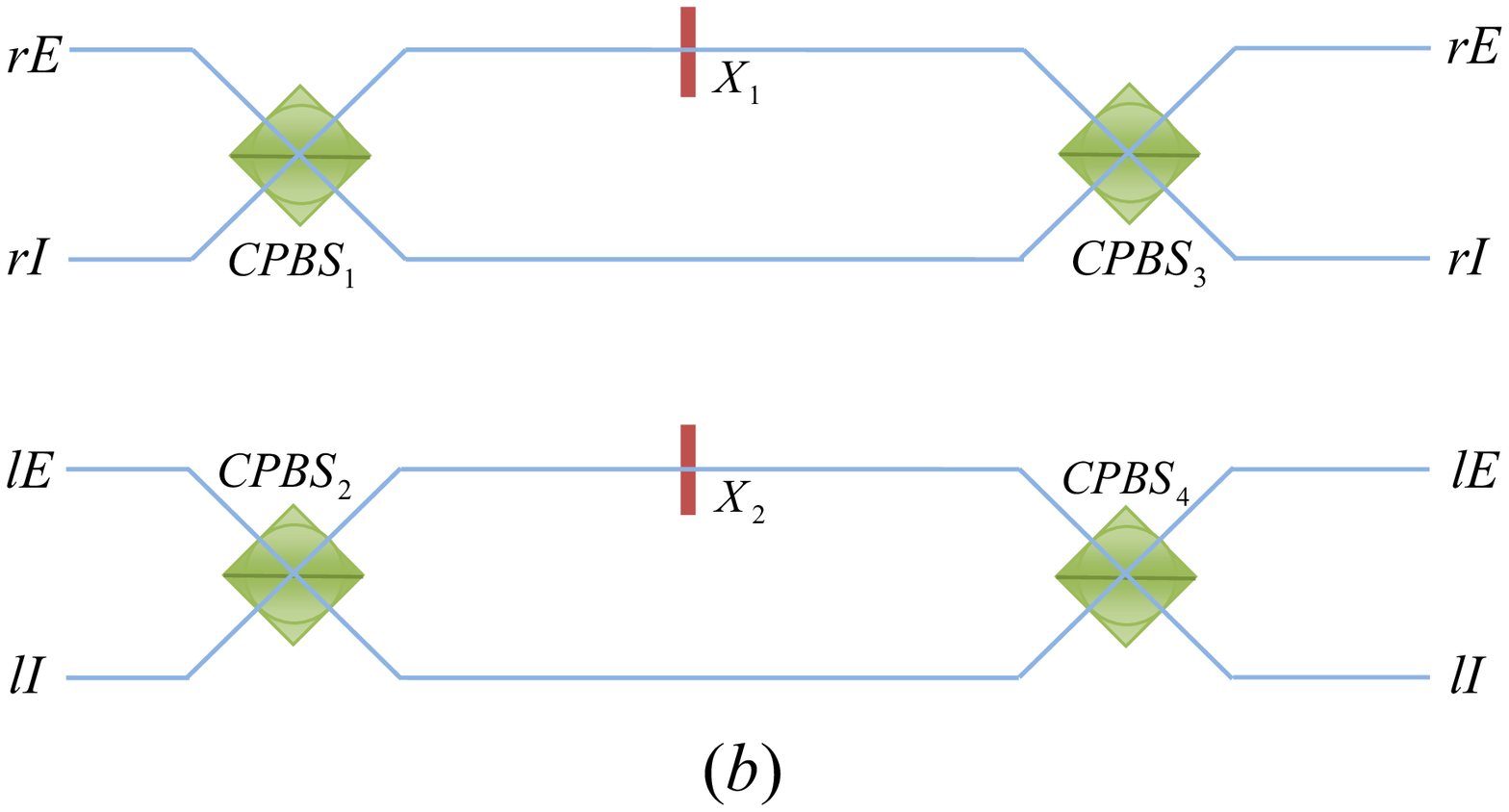}
\caption{(a) Schematic diagram of the SWAP gate between the
polarization state and the spatial-mode state in the first-longitudinal-momentum DOF of one photon. (b) Schematic diagram of
the SWAP gate between the polarization state and the spatial-mode
state in the second-longitudinal-momentum DOF of one photon. $X_{i}$
$(i=1,2)$ is a half-wave plate which performs a polarization
bit-flip operation $\sigma_{x}^{P}=|R\rangle \langle L|+|L\rangle
\langle R|$ on the photon.} \label{fig6}
\end{figure}

\subsection {SWAP gate between the polarization state and the spatial state of one photon}
\label{sec42}

The schematic diagram of our SWAP gate between the polarization
state and the spatial-mode state in the first-longitudinal-momentum
DOF of one photon (P-F-SWAP) is shown in Fig.~\ref{fig6}(a), which
is constructed with linear optical elements such as $CPBS_{i}$
$(i=1,2,3,4)$ and half-wave plate $X_{i}$ $(i=1,2)$. One can inject
photon $A$, which is in the state described as Eq.~(14), into the
quantum circuit shown in Fig.~\ref{fig6}(a). After photon $A$ passes
through the quantum circuit shown in Fig.~\ref{fig6}(a), the state
of photon $A$ is transformed into
\begin{equation}\begin{split}
|\Phi\rangle_{f2}\;=\;&(\gamma_{1}|R\rangle+\delta_{1}|L\rangle)
\otimes  (\alpha_{1}|r\rangle+\beta_{1}|l\rangle) \\
&\otimes (\varepsilon_{1}|E\rangle+\xi_{1}|I\rangle).
\end{split}\end{equation}
Here $|\Phi\rangle_{f2}$ is the objective state of the P-F-SWAP
gate. The schematic diagram of our SWAP gate between the
polarization state and the spatial-mode state in the second
longitudinal-momentum DOF of one photon (P-S-SWAP gate) is shown in
Fig.~\ref{fig6}(b). After photon $A$, whose initial state is
described as Eq.~(14), passes through the circuit shown in
Fig.~\ref{fig6}(b), the state is changed into
\begin{equation}\begin{split}
|\Phi\rangle_{f3}\;=\;&(\varepsilon_{1}|R\rangle+\xi_{1}|L\rangle)
\otimes  (\gamma_{1}|r\rangle+\delta_{1}|l\rangle) \\
& \otimes(\alpha_{1}|E\rangle+\beta_{1}|I\rangle).
\end{split}\end{equation}
Here, $|\Phi\rangle_{f3}$ is just the objective state of the P-S-SWAP gate.

\section{Efficient hyper-EPP for mixed two-photon six-qubit hyperentangled Bell states}
\label{sec5}

In the practical transmission of photons in hyperentangled Bell
states for high-capacity quantum communication, both the bit-flip
error and the phase-flip error will occur on the photon systems.
Although a phase-flip  error cannot be directly purified, it can be
transformed into a bit-flip error using a bilateral local operation
\cite{EPP1,EPP2,EPP3,EPP4,EPP5}. If a bit-flip error purification has
been successfully solved, phase-flip errors also can be solved
perfectly. In this way the two parties in quantum communication, say
Alice and Bob,  can purify a general mixed hyperentangled state.
Below, we only discuss the purification of the two-photon six-qubit
hyperentangled mixed state with bit-flip errors in the three DOFs.

Two identical two-photon six-qubit systems in mixed hyperentangled
Bell states in the polarization DOF and two longitudinal-momentum
DOFs with bit-flip errors can be described as follows:
\begin{equation}\begin{split}
\rho_{AB}\;=\;&[F_{1}|\phi^{+}\rangle_{AB}^{p} \langle \phi^{+}|+(1-F_{1})|\psi^{+}\rangle_{AB}^{p} \langle \psi^{+}|]\\
&\otimes[F_{2}|\phi^{+}\rangle_{AB}^{F} \langle \phi^{+}|+(1-F_{2})|\psi^{+}\rangle_{AB}^{F} \langle \psi^{+}|]\\
&\otimes[F_{3}|\phi^{+}\rangle_{AB}^{S} \langle \phi^{+}|+(1-F_{3})|\psi^{+}\rangle_{AB}^{S} \langle \psi^{+}|], \\
\rho_{CD}\;=\;&[F_{1}|\phi^{+}\rangle_{CD}^{p} \langle \phi^{+}|+(1-F_{1})|\psi^{+}\rangle_{CD}^{p} \langle \psi^{+}|]\\
&\otimes [F_{2}|\phi^{+}\rangle_{CD}^{F} \langle \phi^{+}|+(1-F_{2})|\psi^{+}\rangle_{CD}^{F} \langle \psi^{+}|]\\
&\otimes [F_{3}|\phi^{+}\rangle_{CD}^{S} \langle \phi^{+}|+(1-F_{3})|\psi^{+}\rangle_{CD}^{S} \langle \psi^{+}|].
\end{split}\end{equation}
Here the subscripts $AB$ and $CD$ represent two photon pairs shared
by the two parties in quantum communication , say Alice and Bob.
Alice holds the photons $A$ and $C$, and Bob holds the photons $B$
and $D$. $F_{1}$, $F_{2}$, and $F_{3}$ represent the probabilities
of states $|\phi^{+}\rangle_{AB}^{p}$ $(|\phi^{+}\rangle_{CD}^{p})$,
$|\phi^{+}\rangle_{AB}^{F}$ $(|\phi^{+}\rangle_{CD}^{F})$, and
$|\phi^{+}\rangle_{AB}^{S}$ $(|\phi^{+}\rangle_{CD}^{S})$ in the
mixed states, respectively.

The initial state of the system composed of the two identical
two-photon six-qubit subsystems $ABCD$ can be expressed as
$\rho_{0}=\rho_{AB} \otimes \rho_{CD}$. It can be viewed as a
mixture of maximally hyperentangled pure states. In the polarization
DOF, it is a mixture of the states $|\phi^{+}\rangle_{AB}^{p}
\otimes |\phi^{+}\rangle_{CD}^{p}$, $|\phi^{+}\rangle_{AB}^{p}
\otimes |\psi^{+}\rangle_{CD}^{p}$, $|\psi^{+}\rangle_{AB}^{p}
\otimes |\phi^{+}\rangle_{CD}^{p}$, and $|\psi^{+}\rangle_{AB}^{p}
\otimes |\psi^{+}\rangle_{CD}^{p}$ with the probabilities
$F_{1}^{2}$, $(1-F_{1})F_{1}$, $(1-F_{1})F_{1}$, and
$(1-F_{1})^{2}$, respectively. In the first-longitudinal-momentum
DOF, it is a mixture of the states $|\phi^{+}\rangle_{AB}^{F}
\otimes |\phi^{+}\rangle_{CD}^{F}$, $|\phi^{+}\rangle_{AB}^{F}
\otimes |\psi^{+}\rangle_{CD}^{F}$, $|\psi^{+}\rangle_{AB}^{F}
\otimes |\phi^{+}\rangle_{CD}^{F}$, and $|\psi^{+}\rangle_{AB}^{F}
\otimes |\psi^{+}\rangle_{CD}^{F}$ with the probabilities
$F_{2}^{2}$, $(1-F_{2})F_{2}$, $(1-F_{2})F_{2}$, and
$(1-F_{2})^{2}$, respectively. In the second-longitudinal-momentum
DOF, it is a mixture of the states $|\phi^{+}\rangle_{AB}^{S}
\otimes |\phi^{+}\rangle_{CD}^{S}$, $|\phi^{+}\rangle_{AB}^{S}
\otimes |\psi^{+}\rangle_{CD}^{S}$, $|\psi^{+}\rangle_{AB}^{S}
\otimes |\phi^{+}\rangle_{CD}^{S}$, and $|\psi^{+}\rangle_{AB}^{S}
\otimes |\psi^{+}\rangle_{CD}^{S}$ with the probabilities
$F_{3}^{2}$, $(1-F_{3})F_{3}$, $(1-F_{3})F_{3}$, and
$(1-F_{3})^{2}$, respectively.

Our hyper-EPP for the nonlocal two-photon six-qubit systems in
hyperentangled Bell states with bit-flip errors in the polarization
DOF and the two longitudinal-momentum DOFs can be achieved with two
steps in each round. The first step is completed with polarization
and spatial-mode parity-check QNDs introduced in Sec.~\ref{sec3}.
The second step of our hyper-EPP scheme is completed with the SWAP
gates introduced in Sec.~\ref{sec4}. We discuss the principles of
these two steps in detail as follows.

\begin{figure}[th]%[tpb]
\centering
\includegraphics[width=8cm,angle=0]{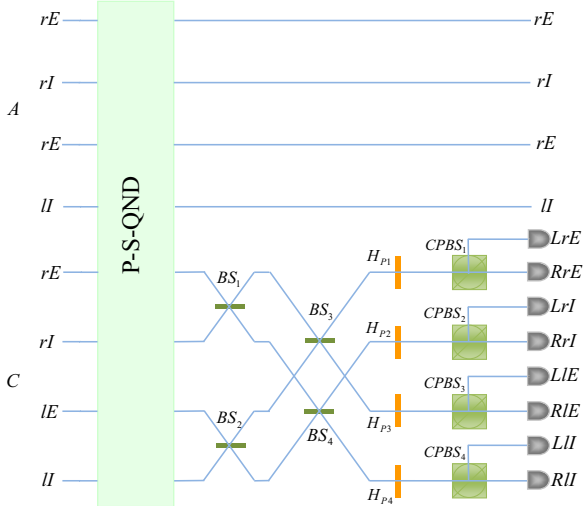}
\caption{Schematic diagram of the first step of our hyper-EPP for
mixed hyperentangled Bell states with bit-flip errors in the
polarization DOF and the two longitudinal-momentum DOFs with
P-S-QNDs. A beam splitter $BS_{i}$ $(i=1,2)$ is used to perform the
Hadamard operation [$|E\rangle \rightarrow
\frac{1}{\sqrt{2}}(|E\rangle+|I\rangle),\;|I\rangle \rightarrow
\frac{1}{\sqrt{2}}(|E\rangle-|I\rangle)$] on the first-longitudinal-momentum DOF of a photon and $BS_{j}$ $(j=3,4)$ is used
to perform the Hadamard operation [$|r\rangle \rightarrow
\frac{1}{\sqrt{2}}(|r\rangle+|l\rangle),\;|l\rangle \rightarrow
\frac{1}{\sqrt{2}}(|r\rangle-|l\rangle)$] on the second-longitudinal-momentum DOF of a photon. $D_{k}$ $(k=LrE, RrE, LrI, RrI, LlE, RlE,
LlI, or RlI)$ is a single-photon detector.}\label{fig7}
\end{figure}

\subsection{The first step of our hyper-EPP with polarization and spatial-mode parity-check QND}

The principle of the first step of our hyper-EPP is shown in
Fig.~\ref{fig7}. Alice performs the P-S-QNDs on photons $AC$ and
performs the Hadamard operations on the polarization DOF and the
spatial-mode DOFs of photon $C$. Bob performs the same operations on
photons $BD$.

First, Alice and Bob perform the P-S-QNDs on the two two-photon
systems $AC$ and $BD$, respectively. Based on the results of the
P-S-QNDs on the identical two-photon systems, the states can be
classified into eight cases. In case (1), the two identical two-photon
systems $AC$ and $BD$ are in the same parity-mode in all of the three
DOFs, including the polarization DOF, the first-longitudinal-momentum
DOF, and the second-longitudinal-momentum DOF. This case
corresponds to the states $|\phi^{+}\rangle_{AB}^{p}
|\phi^{+}\rangle_{CD}^{p}$ or $|\psi^{+}\rangle_{AB}^{p}
|\psi^{+}\rangle_{CD}^{p}$ in the polarization DOF, the states
$|\phi^{+}\rangle_{AB}^{F} |\phi^{+}\rangle_{CD}^{F}$ or
$|\psi^{+}\rangle_{AB}^{F} |\psi^{+}\rangle_{CD}^{F}$ in the first-longitudinal-momentum DOF, and the states $|\phi^{+}\rangle_{AB}^{S}
|\phi^{+}\rangle_{CD}^{S}$ or $|\psi^{+}\rangle_{AB}^{S}
|\psi^{+}\rangle_{CD}^{S}$ in the second-longitudinal-momentum DOF.
The classification of the eight cases is shown in Table~\ref{table1}.

\begin{table}%[htb]
\centering
\caption{The classification based on the results of the P-S-QNDs on photons $AC$ and $BD$.}
\begin{tabular}{cccc}\hline\hline
Case      & DOF               & Parity-mode  & Bell states     \\\hline
          & p                 & Same         & $|\phi^{+}\rangle_{AB}^{p} |\phi^{+}\rangle_{CD}^{p}$ or $|\psi^{+}\rangle_{AB}^{p} |\psi^{+}\rangle_{CD}^{p}$ \\%\cline{2-4}
(1)       & F                 & Same         & $|\phi^{+}\rangle_{AB}^{F} |\phi^{+}\rangle_{CD}^{F}$ or $|\psi^{+}\rangle_{AB}^{F} |\psi^{+}\rangle_{CD}^{F}$ \\%\cline{2-4}
          & S                 & Same         & $|\phi^{+}\rangle_{AB}^{S} |\phi^{+}\rangle_{CD}^{S}$ or $|\psi^{+}\rangle_{AB}^{S} |\psi^{+}\rangle_{CD}^{S}$ \\
          \\%\cline{1-4}
          & p                 & different    & $|\phi^{+}\rangle_{AB}^{p} |\psi^{+}\rangle_{CD}^{p}$ or $|\psi^{+}\rangle_{AB}^{p} |\phi^{+}\rangle_{CD}^{p}$ \\%\cline{2-4}
(2)       & F                 & Different    & $|\phi^{+}\rangle_{AB}^{F} |\psi^{+}\rangle_{CD}^{F}$ or $|\psi^{+}\rangle_{AB}^{F} |\phi^{+}\rangle_{CD}^{F}$ \\%\cline{2-4}
          & S                 & Different    & $|\phi^{+}\rangle_{AB}^{S} |\psi^{+}\rangle_{CD}^{S}$ or $|\psi^{+}\rangle_{AB}^{S} |\phi^{+}\rangle_{CD}^{S}$ \\
          \\%\cline{1-4}
          & p                 & Different    & $|\phi^{+}\rangle_{AB}^{p} |\psi^{+}\rangle_{CD}^{p}$ or $|\psi^{+}\rangle_{AB}^{p} |\phi^{+}\rangle_{CD}^{p}$ \\%\cline{2-4}
(3)       & F                 & Same         & $|\phi^{+}\rangle_{AB}^{F} |\phi^{+}\rangle_{CD}^{F}$ or $|\psi^{+}\rangle_{AB}^{F} |\psi^{+}\rangle_{CD}^{F}$ \\%\cline{2-4}
          & S                 & Same         & $|\phi^{+}\rangle_{AB}^{S} |\phi^{+}\rangle_{CD}^{S}$ or $|\psi^{+}\rangle_{AB}^{S} |\psi^{+}\rangle_{CD}^{S}$ \\
          \\%\cline{1-4}
          & p                 & Same         & $|\phi^{+}\rangle_{AB}^{p} |\phi^{+}\rangle_{CD}^{p}$ or $|\psi^{+}\rangle_{AB}^{p} |\psi^{+}\rangle_{CD}^{p}$ \\%\cline{2-4}
(4)       & F                 & Different    & $|\phi^{+}\rangle_{AB}^{F} |\psi^{+}\rangle_{CD}^{F}$ or $|\psi^{+}\rangle_{AB}^{F} |\phi^{+}\rangle_{CD}^{F}$ \\%\cline{2-4}
          & S                 & Same         & $|\phi^{+}\rangle_{AB}^{S} |\phi^{+}\rangle_{CD}^{S}$ or $|\psi^{+}\rangle_{AB}^{S} |\psi^{+}\rangle_{CD}^{S}$ \\
          \\%\cline{1-4}
          & p                 & Same         & $|\phi^{+}\rangle_{AB}^{p} |\phi^{+}\rangle_{CD}^{p}$ or $|\psi^{+}\rangle_{AB}^{p} |\psi^{+}\rangle_{CD}^{p}$ \\%\cline{2-4}
(5)       & F                 & Same         & $|\phi^{+}\rangle_{AB}^{F} |\phi^{+}\rangle_{CD}^{F}$ or $|\psi^{+}\rangle_{AB}^{F} |\psi^{+}\rangle_{CD}^{F}$ \\%\cline{2-4}
          & S                 & Different    & $|\phi^{+}\rangle_{AB}^{S} |\psi^{+}\rangle_{CD}^{S}$ or $|\psi^{+}\rangle_{AB}^{S} |\phi^{+}\rangle_{CD}^{S}$ \\
          \\%\cline{1-4}
          & p                 & Different    & $|\phi^{+}\rangle_{AB}^{p} |\psi^{+}\rangle_{CD}^{p}$ or $|\psi^{+}\rangle_{AB}^{p} |\phi^{+}\rangle_{CD}^{p}$ \\%\cline{2-4}
(6)       & F                 & Different    & $|\phi^{+}\rangle_{AB}^{F} |\psi^{+}\rangle_{CD}^{F}$ or $|\psi^{+}\rangle_{AB}^{F} |\phi^{+}\rangle_{CD}^{F}$ \\%\cline{2-4}
          & S                 & Same         & $|\phi^{+}\rangle_{AB}^{S} |\phi^{+}\rangle_{CD}^{S}$ or $|\psi^{+}\rangle_{AB}^{S} |\psi^{+}\rangle_{CD}^{S}$ \\
          \\%\cline{1-4}
          & p                 & Different    & $|\phi^{+}\rangle_{AB}^{p} |\psi^{+}\rangle_{CD}^{p}$ or $|\psi^{+}\rangle_{AB}^{p} |\phi^{+}\rangle_{CD}^{p}$ \\%\cline{2-4}
(7)       & F                 & Same         & $|\phi^{+}\rangle_{AB}^{F} |\phi^{+}\rangle_{CD}^{F}$ or $|\psi^{+}\rangle_{AB}^{F} |\psi^{+}\rangle_{CD}^{F}$ \\%\cline{2-4}
          & S                 & Different    & $|\phi^{+}\rangle_{AB}^{S} |\psi^{+}\rangle_{CD}^{S}$ or $|\psi^{+}\rangle_{AB}^{S} |\phi^{+}\rangle_{CD}^{S}$ \\
          \\%\cline{1-4}
          & p                 & Same         & $|\phi^{+}\rangle_{AB}^{p} |\phi^{+}\rangle_{CD}^{p}$ or $|\psi^{+}\rangle_{AB}^{p} |\psi^{+}\rangle_{CD}^{p}$ \\%\cline{2-4}
(8)       & F                 & Different    & $|\phi^{+}\rangle_{AB}^{F} |\psi^{+}\rangle_{CD}^{F}$ or $|\psi^{+}\rangle_{AB}^{F} |\phi^{+}\rangle_{CD}^{F}$ \\%\cline{2-4}
          & S                 & Different    & $|\phi^{+}\rangle_{AB}^{S} |\psi^{+}\rangle_{CD}^{S}$ or $|\psi^{+}\rangle_{AB}^{S} |\phi^{+}\rangle_{CD}^{S}$ \\ \hline\hline %\cline{1-4}\cline{1-4}
\end{tabular}\label{table1}
\end{table}

Now, let us discuss case (1), in which the parity modes of $AC$
and $BD$ are same in the polarization DOF, the first-longitudinal-momentum DOF, and the second-longitudinal-momentum DOF, to bring to
light of the principle of the first step in our hyper-EPP. Case
(2), in which the parity modes of $AC$ and $BD$ are different in the
polarization DOF, the first-longitudinal-momentum DOF, and the
second-longitudinal-momentum DOF, is discussed in the Appendix. The other cases are
similar to these two cases with a little modification. When $AC$
and $BD$ are both in the even-parity modes in the polarization DOF,
the first-longitudinal-momentum DOF, and the second longitudinal
momentum DOF, Alice and Bob obtain the states
 \begin{equation}\begin{split}
&|\Phi_{1}\rangle^{p}=\frac{1}{\sqrt{2}}(|RRRR\rangle+|LLLL\rangle)_{ABCD},\\
&|\Phi_{2}\rangle^{p}=\frac{1}{\sqrt{2}}(|RLRL\rangle+|LRLR\rangle)_{ABCD},\\
&|\Phi_{1}\rangle^{F}=\frac{1}{\sqrt{2}}(|rrrr\rangle+|llll\rangle)_{ABCD},\\
&|\Phi_{2}\rangle^{F}=\frac{1}{\sqrt{2}}(|rlrl\rangle+|lrlr\rangle)_{ABCD},\\
&|\Phi_{1}\rangle^{S}=\frac{1}{\sqrt{2}}(|EEEE\rangle+|IIII\rangle)_{ABCD},\\
&|\Phi_{2}\rangle^{S}=\frac{1}{\sqrt{2}}(|EIEI\rangle+|IEIE\rangle)_{ABCD}.
 \end{split}\end{equation}
On the contrary, when $AC$ and $BD$ are both in the odd-parity modes
in the polarization DOF,  the first-longitudinal-momentum DOF, and
the second-longitudinal-momentum DOF, Alice and Bob obtain the
states
\begin{equation}\begin{split}
&|\Phi_{3}\rangle^{p}=\frac{1}{\sqrt{2}}(|RRLL\rangle+|LLRR\rangle)_{ABCD},\\
& |\Phi_{4}\rangle^{p}=\frac{1}{\sqrt{2}}(|RLLR\rangle+|LRRL\rangle)_{ABCD},\\
&|\Phi_{3}\rangle^{F}=\frac{1}{\sqrt{2}}(|rrll\rangle+|llrr\rangle)_{ABCD},\\
&|\Phi_{4}\rangle^{F}=\frac{1}{\sqrt{2}}(|rllr\rangle+|lrrl\rangle)_{ABCD},\\
& |\Phi_{3}\rangle^{S}=\frac{1}{\sqrt{2}}(|EEII\rangle+|IIEE\rangle)_{ABCD},\\
&|\Phi_{4}\rangle^{S}=\frac{1}{\sqrt{2}}(|EIIE\rangle+|IEEI\rangle)_{ABCD}.
 \end{split}\end{equation}
The states $|\Phi_{3}\rangle^{p}$ and $|\Phi_{4}\rangle^{p}$ can be
transformed into $|\Phi_{1}\rangle^{p}$ and $|\Phi_{2}\rangle^{p}$
by performing the bit-flip operations on photons $CD$ in the
polarization DOF, respectively. Similarly, $|\Phi_{3}\rangle^{F}$
and $|\Phi_{4}\rangle^{F}$ can be transformed into
$|\Phi_{1}\rangle^{F}$ and $|\Phi_{2}\rangle^{F}$ by the bit-flip
operations on photons $CD$ in the first-longitudinal-momentum DOF,
respectively. $|\Phi_{3}\rangle^{S}$ and $|\Phi_{4}\rangle^{S}$ can
also be transformed into $|\Phi_{1}\rangle^{S}$ and
$|\Phi_{2}\rangle^{S}$ by the bit-flip operations on photons $CD$ in
the second-longitudinal-momentum DOF, respectively.

Next, Alice and Bob perform Hadamard operations on photons $C$ and
$D$ in the polarization and the two longitudinal-momentum DOFs,
respectively. In the polarization DOF, the states
$|\Phi_{1}\rangle^{p}$ and $|\Phi_{2}\rangle^{p}$ are transformed
into the states $|\Phi_{1'}\rangle^{p}$ and $|\Phi_{2'}\rangle^{p}$,
respectively. In the first-longitudinal-momentum DOF, the states
$|\Phi_{1}\rangle^{F}$ and $|\Phi_{2}\rangle^{F}$ are transformed
into $|\Phi_{1}'\rangle^{F}$ and $|\Phi_{2}'\rangle^{F}$, respectively.
In the second-longitudinal-momentum DOF, the states
$|\Phi_{1}\rangle^{S}$ and $|\Phi_{2}\rangle^{S}$ are transformed
into $|\Phi_{1}'\rangle^{S}$ and $|\Phi_{2}'\rangle^{S}$, respectively.
Here
\begin{equation}\begin{split}
|\Phi'_{1}\rangle^{p}\;=\;&\frac{1}{2\sqrt{2}}[(|RR\rangle+|LL\rangle)_{AB}
(|RR\rangle+|LL\rangle)_{CD}\\
&+(|RR\rangle-|LL\rangle)_{AB}(|RL\rangle+|LR\rangle)_{CD}],\\
|\Phi'_{2}\rangle^{p}\;=\;&\frac{1}{2\sqrt{2}}[(|RL\rangle+|LR\rangle)_{AB}
(|RR\rangle-|LL\rangle)_{CD}\\
&+(-|RL\rangle+|LR\rangle)_{AB}(|RL\rangle-|LR\rangle)_{CD}], \\
|\Phi'_{1}\rangle^{F}\;=\;&\frac{1}{2\sqrt{2}}[(|rr\rangle+|ll\rangle)_{AB}
(|rr\rangle+|ll\rangle)_{CD}\\
&+(|rr\rangle-|ll\rangle)_{AB}(|rl\rangle+|lr\rangle)_{CD}],\\
|\Phi'_{2}\rangle^{F}\;=\;&\frac{1}{2\sqrt{2}}[(|rl\rangle+|lr\rangle)_{AB}
(|rr\rangle-|ll\rangle)_{CD}\\
&+(-|rl\rangle+|lr\rangle)_{AB}(|rl\rangle-|lr\rangle)_{CD}],\\
|\Phi'_{1}\rangle^{S}\;=\;&\frac{1}{2\sqrt{2}}[(|EE\rangle+|II\rangle)_{AB}
(|EE\rangle+|II\rangle)_{CD}\\
&+(|EE\rangle-|II\rangle)_{AB}(|EI\rangle+|IE\rangle)_{CD}],\\
|\Phi'_{2}\rangle^{S}\;=\;&\frac{1}{2\sqrt{2}}[(|EI\rangle+|IE\rangle)_{AB}
(|EE\rangle-|II\rangle)_{CD}\\
&+(-|EI\rangle+|IE\rangle)_{AB}(|EI\rangle-|IE\rangle)_{CD}].
\end{split}\end{equation}

Finally, the photons $C$ and $D$ are detected by single-photon
detectors, respectively. If the photons $CD$ are in the even-parity
mode in the polarization DOF (the first-longitudinal-momentum DOF or the
second-longitudinal-momentum DOF), nothing is needed to be done. If the
photons $CD$ are in the odd-parity mode in the polarization DOF (the
first-longitudinal-momentum DOF or the second-longitudinal-momentum
DOF), the phase-flip operation $\sigma_{z}^{p}=|R\rangle\langle
R|-|L\rangle \langle L|$ ($\sigma_{z}^{F}=|r\rangle\langle
r|-|l\rangle \langle l|$ or $\sigma_{z}^{S}=|E\rangle\langle
E|-|I\rangle \langle I|$) is operated on the photon $B$.

\begin{table*}
\caption{The probabilities of different states corresponding to the
eight cases, respectively.}
\begin{tabular}{ccccccccccccccccccc}\hline\hline
\multirow{2}*{Case}     &
\multicolumn{16}{c}{Probability}\\\cline{4-19}
       &&    &$|\phi^{+}\rangle_{AB}^{p}$ &&  &$|\psi^{+}\rangle_{AB}^{p}$ && &$|\phi^{+}\rangle_{AB}^{F}$ &&  &$|\psi^{+}\rangle_{AB}^{F}$ && &$|\phi^{+}\rangle_{AB}^{S}$ &&  &$|\psi^{+}\rangle_{AB}^{S}$ \\\hline
(1)  &&      &$F_{1}^{2}$   &&       &$(1-F_{1})^{2}$   &&      &$F_{2}^{2}$  &&           &$(1-F_{2})^{2}$   &&      &$F_{3}^{2}$      &&       &$(1-F_{3})^{2}$        \\%\hline
(2)  &&      &$F_{1}(1-F_{1})$   &&     &$F_{1}(1-F_{1})$    &&    &$F_{2}(1-F_{2})$  &&      &$F_{2}(1-F_{2})$   &&     &$F_{3}(1-F_{3})$   &&     &$F_{3}(1-F_{3})$        \\%\hline
(3)  &&      &$F_{1}(1-F_{1})$   &&     &$F_{1}(1-F_{1})$   &&     &$F_{2}^{2}$  &&           &$(1-F_{2})^{2}$    &&     &$(1-F_{3})^{2}$    &&     &$(1-F_{3})^{2}$        \\%\hline
(4)  &&      &$F_{1}^{2}$     &&        &$(1-F_{1})^{2}$    &&     &$F_{2}(1-F_{2})$  &&      &$F_{2}(1-F_{2})$   &&     &$F_{3}^{2}$      &&       &$(1-F_{3})^{2}$       \\%\hline
(5)  &&      &$F_{1}^{2}$    &&         &$(1-F_{1})^{2}$    &&     &$F_{2}^{2}$  &&           &$(1-F_{2})^{2}$   &&      &$F_{3}(1-F_{3})$   &&     &$F_{3}(1-F_{3})$       \\%\hline
(6)  &&      &$F_{1}(1-F_{1})$   &&     &$F_{1}(1-F_{1})$   &&     &$F_{2}(1-F_{2})$  &&      &$F_{2}(1-F_{2})$   &&     &$F_{3}^{2}$     &&        &$(1-F_{3})^{2}$        \\%\hline
(7)  &&      &$F_{1}(1-F_{1})$   &&     &$F_{1}(1-F_{1})$   &&     &$F_{2}^{2}$  &&           &$(1-F_{2})^{2}$   &&      &$F_{3}(1-F_{3})$   &&     &$F_{3}(1-F_{3})$       \\%\hline
(8)  &&      &$F_{1}^{2}$     &&        &$(1-F_{1})^{2}$    &&     &$F_{2}(1-F_{2})$  &&       &$F_{2}(1-F_{2})$   &&     &$F_{3}(1-F_{3})$   &&     &$F_{3}(1-F_{3})$        \\ \hline\hline
\end{tabular}\label{table2}
\end{table*}

After the first step of our hyper-EPP, Alice and Bob obtain the
states $|\phi^{+}\rangle^{p\,(F,S)}$ and
$|\psi^{+}\rangle^{p\,(F,S)}$ with different probabilities
corresponding to the eight cases, respectively, which are shown in
Table ~\ref{table2}. If the two two-photon systems $ABCD$ are
projected into the states in case (1), the hyper-EPP of the system
$AB$ is completed. If the two two-photon systems $ABCD$ are
projected into case (2), Alice and Bob will discard the two
photons. Otherwise, the second step of hyper-EPP with SWAP gates is
required if the two two-photon systems $ABCD$ are projected into
case ($i$) ($i=3\sim8$).

\subsection{The second step of our  hyper-EPP with SWAP gates}

If the two two-photon systems $AB$ and $CD$ are projected into the
case (6), the fidelities of the Bell states of the photon pair $AB$
in the polarization and the first-longitudinal-momentum DOFs are
lower than the initial ones and the fidelity in the second-longitudinal-momentum DOF is higher than the initial one. Alice and Bob seek
another two two-photon systems $A'B'$ and $C'D'$ projected into the
case (5), whose fidelity is lower than the initial one in the
second-longitudinal-momentum DOF and higher than the initial one in
the polarization and the first-longitudinal-momentum DOFs,
respectively. The photons $A$, $C$, $A'$, and $C'$ belong to Alice
and $B$, $D$, $B'$, and $D'$ belong to Bob. We discuss the result of
the second step when the states of the photons $AB$ and $A'B'$ are
in
$|\Phi\rangle_{AB0}=|\psi^{+}\rangle^{p}|\phi^{+}\rangle^{F}|\phi^{+}\rangle^{S}$
and
$|\Phi\rangle_{A'B'0}=|\phi^{+}\rangle^{p}|\psi^{+}\rangle^{F}|\psi^{+}\rangle^{S}$,
respectively, as an example.

First, Alice and Bob swap the polarization states of the two-photon system $AB$ and the states of the system $A'B'$ by using
the P-P-SWAP gate shown in Fig.~\ref{fig5}. The state of the system
composed of four-photon $ABA'B'$ and two NV centers are
transformed from
$|\Phi\rangle_{0}=|\Phi\rangle_{AB0}|\Phi\rangle_{A'B'0}
|\phi_{+}\rangle_{a}|\phi_{+}\rangle_{b}$ to the state
\begin{equation}\begin{split}
|\Phi\rangle
\;=\;&\frac{1}{4}[-(|RRRL\rangle+|LLRL\rangle\\
&+|RRLR\rangle+|LLLR\rangle)|+1\rangle_{a}|+1\rangle_{b}\\
&+(|RRRL\rangle-|LLRL\rangle\\
&-|RRLR\rangle+|LLLR\rangle)|+1\rangle_{a}|-1\rangle_{b} \\
&-(|RRRL\rangle-|LLRL\rangle\\
&-|RRLR\rangle+|LLLR\rangle)|-1\rangle_{a}|+1\rangle_{b} \\
&+(|RRRL\rangle+|LLRL\rangle\\
&+|RRLR\rangle+|LLLR\rangle)|-1\rangle_{a}|-1\rangle_{b} ]\\
&\otimes
|\phi^{+}\rangle^{F}_{AB}|\phi^{+}\rangle^{S}_{AB}|\psi^{+}\rangle^{F}_{A'B'}|\psi^{+}\rangle^{S}_{A'B'}.
\end{split}\end{equation}
If the state of two NV centers $a$ and $b$ belonging to Alice and
Bob is  $|+1\rangle_{a}|+1\rangle_{b}$ or $|-1\rangle_{a}|-1\rangle_{b}$, Alice and Bob do nothing.  If the state of
the NV centers is  $|+1\rangle_{a}|-1\rangle_{b}$ or $|-1\rangle_{a}|+1\rangle_{b}$, Alice and Bob
perform a $\sigma_{z}^{P}=|R\rangle\langle R|- |L\rangle\langle L|$ operation on photons $A'$ and $B$, respectively. Thus, Alice and Bob obtain the states
\begin{equation}\begin{split}
&|\Phi\rangle_{AB1}=|\phi^{+}\rangle^{p}|\phi^{+}\rangle^{F}|\phi^{+}\rangle^{S}, \\
&|\Phi\rangle_{A'B'1}=|\psi^{+}\rangle^{p}|\psi^{+}\rangle^{F}|\psi^{+}\rangle^{S}.
\end{split}\end{equation}

Second, Alice and Bob swap the polarization states and the first-longitudinal-momentum states of photons $A$, $B$, $A'$, and $B'$,
respectively. The states of the two two-photon systems are transformed into the states
\begin{equation}\begin{split}
&|\Phi\rangle_{AB2}=|\phi^{+}\rangle^{p}|\phi^{+}\rangle^{F}|\phi^{+}\rangle^{S}, \\
&|\Phi\rangle_{A'B'2}=|\psi^{+}\rangle^{p}|\psi^{+}\rangle^{F}|\psi^{+}\rangle^{S}.
\end{split}\end{equation}

Third, Alice and Bob swap the polarization states of
the two systems $AB$ and $A'B'$ again and obtain the states
\begin{equation}\begin{split}
&|\Phi\rangle_{AB3}=|\psi^{+}\rangle^{p}|\phi^{+}\rangle^{F}|\phi^{+}\rangle^{S}, \\
&|\Phi\rangle_{A'B'3}=|\phi^{+}\rangle^{p}|\psi^{+}\rangle^{F}|\psi^{+}\rangle^{S}.
\end{split}
\end{equation}

Finally, Alice and Bob swap the polarization states and the first-longitudinal-momentum states of
photons $A$, $B$, $A'$, and $B'$, respectively. The states of the
two  systems  are changed into
\begin{equation}\begin{split}
&|\Phi\rangle_{AB4}=|\phi^{+}\rangle^{p}|\psi^{+}\rangle^{F}|\phi^{+}\rangle^{S}\\
&|\Phi\rangle_{A'B'4}=|\psi^{+}\rangle^{p}|\phi^{+}\rangle^{F}|\psi^{+}\rangle^{S}.
\end{split}\end{equation}
Here $|\Phi\rangle_{AB4}$ and $|\Phi\rangle_{A'B'4}$ are the final states of this case in the second step of hyper-EPP.

If the two two-photon systems $AB$ and $CD$ are projected into the
states in case (3), (4), (5), (7), or (8), the second step of the
hyper-EPP can be completed, as  discussed in the Appendix.

\begin{figure}[th]%[tpb]
\centering
\includegraphics[width=6cm,angle=0]{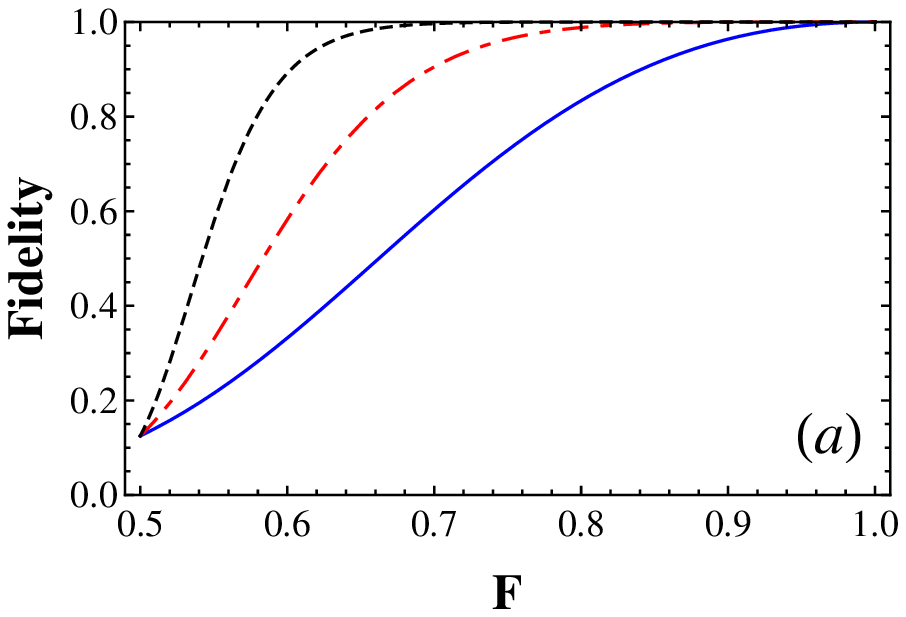} \hspace{30pt}
\includegraphics[width=6cm,angle=0]{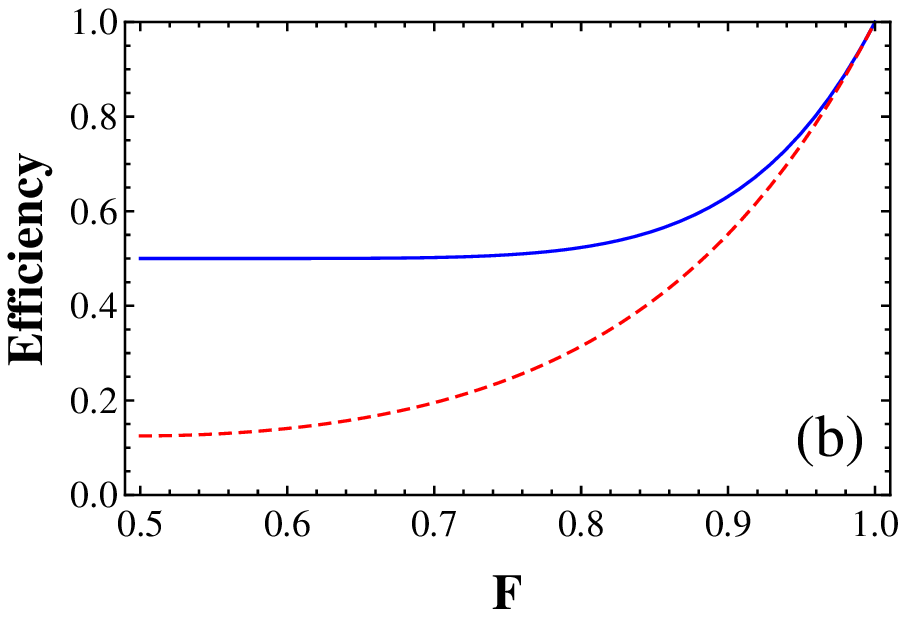}
\caption{(a) The fidelities of our hyper-EPP for mixed hyperentangled Bell states in three DOFs.
The solid line, dash-dotted line, and dashed line represent the fidelities of the iteration times $n=1$, $2$, and $3$,
respectively, for the cases with $F_{1}=F_{2}=F_{3}=F$. (b) The efficiencies of the of
our hyper-EPP for mixed hyperentangled Bell states in three DOFs. The dashed line and
the solid line represent the efficiencies of the first step and the two steps in the first round,
respectively for the cases with $F_{1}=F_{2}=F_{3}=F$. } \label{fig8}
\end{figure}

After these two steps, the first round of our hyper-EPP
process is completed and the state of the two-photon system $AB$ is changed to be
\begin{equation}\begin{split}
\rho'_{AB}=&[F'_{1}|\phi^{+}\rangle_{AB}^{p} \langle \phi^{+}|+(1-F'_{1})|\psi^{+}\rangle_{AB}^{p} \langle \psi^{+}|]\\
&\otimes[F'_{2}|\phi^{+}\rangle_{AB}^{F} \langle \phi^{+}|+(1-F'_{2})|\psi^{+}\rangle_{AB}^{F} \langle \psi^{+}|]\\
&\otimes[F'_{3}|\phi^{+}\rangle_{AB}^{S} \langle \phi^{+}|+(1-F'_{3})|\psi^{+}\rangle_{AB}^{S} \langle \psi^{+}|].
\end{split}\end{equation}
Here $F'_{1}=\frac{F_{1}^{2}}{F_{1}^{2}+(1-F_{1})^{2}}$,
$F'_{2}=\frac{F_{2}^{2}}{F_{2}^{2}+(1-F_{2})^{2}}$, and
$F'_{3}=\frac{F_{3}^{2}}{F_{3}^{2}+(1-F_{3})^{2}}$. The fidelity of
the finale state is $F'=F'_{1} F'_{2} F'_{3}$ which is
higher than the initial fidelity $F=F_{1}F_{2} F_{3}$
when $F_{i}>1/2$ $(i=1,2,3)$. By iterating our
hyper-EPP process, the fidelity of the state of the two-photon
system can be improved dramatically. The fidelities of the final
states are shown in Fig.~\ref{fig8}(a) with iterating $n=1,2,3$ times
for the case with $F_{1}=F_{2}=F_{3}=F$.

The efficiency of an EPP is defined as the probability of obtaining
a high-fidelity entangled photon system from a pair of photon
systems transmitted over a noisy channel without photon loss. The
efficiency of our hyper-EPP after the first step in the first round
is
\begin{equation}\begin{split}
Y_{1}=[F_{1}^{2}+(1-F_{1})^{2}][F_{2}^{2}+(1-F_{2})^{2}][F_{3}^{2}+(1-F_{3})^{2}].
\end{split}\end{equation}

After the second step, the efficiency of the first round of the
hyper-EPP process is
\begin{equation}\begin{split}
Y_{2}=&[F_{1}^{2}+(1-F_{1})^{2}][F_{2}^{2}+(1-F_{2})^{2}][F_{3}^{2}+(1-F_{3})^{2}] \\
&+min\{([2F_{1}(1-F_{1})][F_{2}^{2}+(1-F_{2})^{2}]\\
&\;\;\;\;\;\;\;\;\;\;\;\;[F_{3}^{2}+(1-F_{3})^{2}]),([F_{1}^{2}+(1-F_{1})^{2}]\\
&\;\;\;\;\;\;\;\;\;\;\;\;[2F_{2}(1-F_{2})][2F_{3}(1-F_{3})])\}\\
&+min\{([F_{1}^{2}+(1-F_{1})^{2}][2F_{2}(1-F_{2})]\\
&\;\;\;\;\;\;\;\;\;\;\;\;[F_{3}^{2}+(1-F_{3})^{2}]),([2F_{1}(1-F_{1})]\\
&\;\;\;\;\;\;\;\;\;\;\;\;[F_{2}^{2}+(1-F_{2})^{2}][2F_{3}(1-F_{3})])\}\\
&+min\{([F_{1}^{2}+(1-F_{1})^{2}][F_{2}^{2}+(1-F_{2})^{2}]\\
&\;\;\;\;\;\;\;\;\;\;\;\;[2F_{3}(1-F_{3})]),([2F_{1}(1-F_{1})]\\
&\;\;\;\;\;\;\;\;\;\;\;\;[2F_{2}(1-F_{2})][F_{3}^{2}+(1-F_{3})^{2}])\}.
\end{split}\end{equation}

The efficiencies $Y_{1}$ and $Y_{2}$ of our hyper-EPP for the case
with $F_{1}=F_{2}=F_{3}=F$ are shown in Fig.~\ref{fig8}(b),
respectively. Obviously, the efficiency of our hyper-EPP is improved
largely with the second step for purification.

\section{Expansion}
\label{sec6}

We present a hyper-EPP for the nonlocal two-photon systems in the
polarization and two longitudinal-momentum DOFs by two steps. The
first step is completed by the P-S-QNDs, Hadamard operations, and
some detections. After the first step, we can obtain the information
of the probability of each Bell state in different DOFs. If the
probabilities of the three even-parity Bell states in the three DOFs
are higher than the initial ones [corresponding to the case (1) in
the first step], the second step is not need. If the probabilities
of the even-parity Bell states in the three DOFs are all lower than
the initial ones [corresponding to the case (2) in the first step],
the two-photon system is discarded. Otherwise [corresponding to the
case (3), (4), (5), (6), (7) or (8)], the second step is
required. The second step is completed by using the combination of
the P-P-SWAP gate, the P-F-SWAP gate, and the P-S-SWAP gate. The purpose
of the second step is to improve the probability of the even-parity
Bell state in one or two DOFs by costing some other two-photon
systems. For example, if the two two-photon systems $ABCD$ are
projected into the states in case (4), the probability of the
even-parity Bell state in the first-longitudinal-momentum DOF is
lower than the initial one. We seek another two two-photon systems
$A'B'C'D'$ whose probability of even-parity Bell states in the first-longitudinal-momentum DOF is higher than the initial one and the
fidelities of the Bell states in another two DOFs are lower than the
initial ones. Thus, with the P-P-SWAP gate and the P-F-SWAP gate, the
fidelity of the two-photon system $AB$ is improved. If we can construct
the SWAP gate between the polarization state and arbitrary another
longitudinal-momentum state without affecting the states in other
DOFs, the second step can be used to purify the two-photon systems in
the polarization DOF and multiple longitudinal-momentum DOFs. That
is to say, using SWAP gates is a universal method for purifying the
two-photon systems in the polarization DOF and multiple longitudinal-momentum DOFs.

\begin{figure}[th]%[tpb]
\centering
\includegraphics[width=8cm,angle=0]{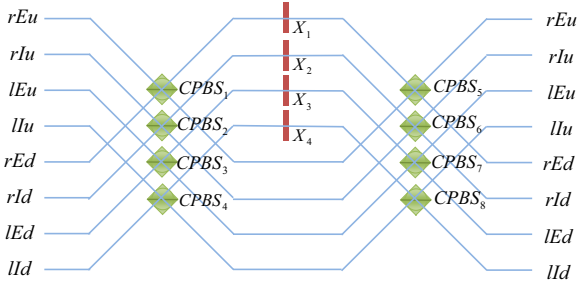}
\caption{Schematic diagram of the SWAP gate between the polarization
state and the spatial-mode state in the third longitudinal-momentum
DOF of one photon.} \label{fig9}
\end{figure}

We construct a SWAP gate between the polarization state and the
state in the third-longitudinal-momentum DOF of a photon (P-T-SWAP
gate) of being simultaneously entangled in the polarization DOF and
three longitudinal-momentum DOFs as an example. The initial state of
a photon being simultaneously entangled with other photons in the polarization DOF and
three longitudinal-momentum DOFs can be expressed as
\begin{equation}\begin{split}
|\Phi\rangle=&(a_{1}|R\rangle+b_{1}|L\rangle)  \otimes (a_{2}|r\rangle+b_{2}|l\rangle) \\
 &\otimes (a_{3}|E\rangle+b_{3}|I\rangle)  \otimes (a_{4}|u\rangle+b_{4}|d\rangle).
\end{split}\end{equation}
Here $(a_{4}|u\rangle+b_{4}|d\rangle)$ is the state of the photon
$A$ being entangled with other photons in the third-longitudinal-momentum DOF ($a_4$ and $b_4$ include not only the parameters for photon $A$ but also the states for other photons, so do $a_i$ and $b_i$). $u$ and $d$ represent
the upper and the down spatial-mode states of the photon,
respectively. The schematic diagram of the P-T-SWAP gate, which is
constructed with some linear optical elements, is shown in
Fig.~\ref{fig9}. After the photon $A$ passes through the quantum
circuit, its state  is transformed into
\begin{equation}\begin{split}
|\Phi'\rangle=&(a_{4}|R\rangle+b_{4}|L\rangle)  \otimes (a_{2}|r\rangle+b_{2}|l\rangle) \\
 &\otimes (a_{3}|E\rangle+b_{3}|I\rangle)  \otimes (a_{1}|u\rangle+b_{1}|d\rangle).
\end{split}\end{equation}
Here $|\Phi'\rangle$ is the objective state of the P-T-SWAP gate.
After the first step of the hyper-EPP, if the probability of the
even-parity Bell state of the system $ABCD$ in the third
longitudinal-momentum DOF is lower than the initial one, we can seek
another system $A'B'C'D'$ whose probability of the even-parity Bell
state in this DOF is higher. By using the P-P-SWAP gate and the P-T-SWAP gate,
the second step of our hyper-EPP can be completed to improve the efficiency of purification largely.

\begin{figure}[th]%[tpb]
\centering
\includegraphics[width=6cm,angle=0]{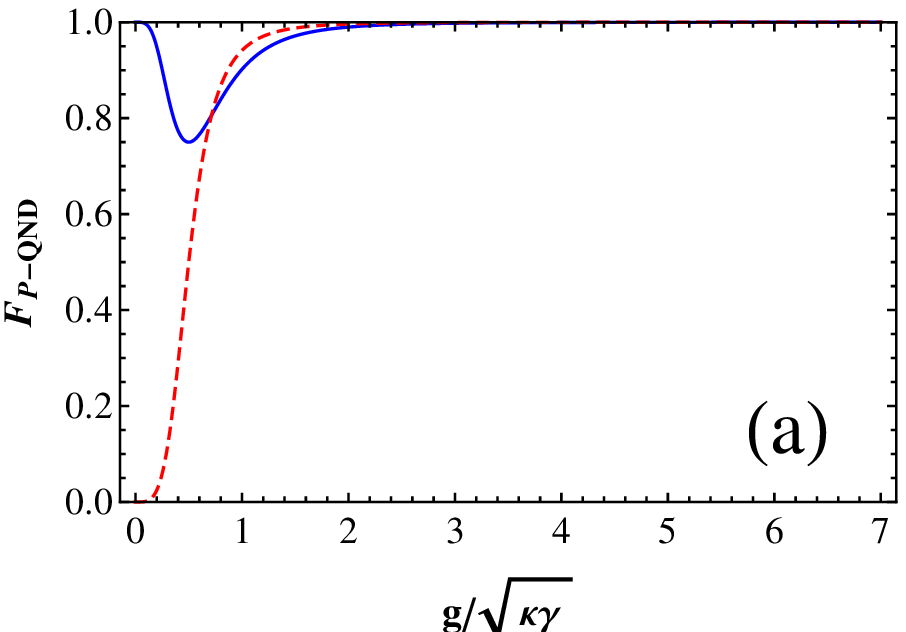} \hspace{30pt}
\includegraphics[width=6cm,angle=0]{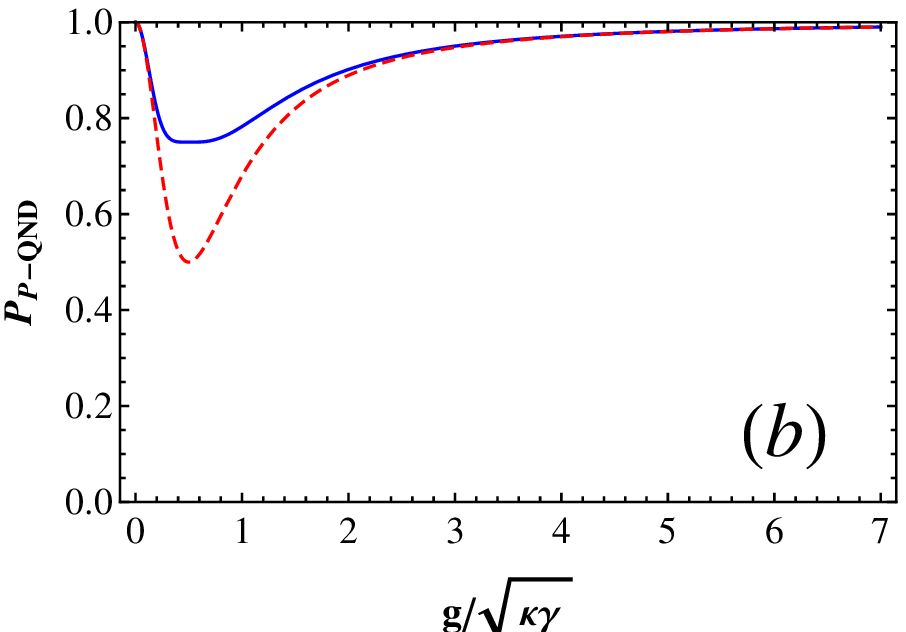}
\caption{(a) The fidelity of the polarization-parity check QND. (b) The efficiency of the polarization-parity check QND.
The solid line represents $F_{P1}$ $(\eta_{P1})$ for $|\phi^{\pm}\rangle^{p} |\phi^{\pm}\rangle^{F} |\phi^{\pm}\rangle^{S}$,
$|\phi^{\pm}\rangle^{p} |\phi^{\pm}\rangle^{F} |\psi^{\pm}\rangle^{S}$, $|\phi^{\pm}\rangle^{p} |\psi^{\pm}\rangle^{F} |\phi^{\pm}\rangle^{S}$,
and $|\phi^{\pm}\rangle^{p} |\psi^{\pm}\rangle^{F} |\psi^{\pm}\rangle^{S}$. The dashed line represents $F_{P2}$ $(\eta_{P2})$
for $|\psi^{\pm}\rangle^{p} |\phi^{\pm}\rangle^{F} |\phi^{\pm}\rangle^{S}$, $|\psi^{\pm}\rangle^{p} |\phi^{\pm}\rangle^{F} |\psi^{\pm}\rangle^{S}$,
$|\psi^{\pm}\rangle^{p} |\psi^{\pm}\rangle^{F} |\phi^{\pm}\rangle^{S}$, and $|\psi^{\pm}\rangle^{p} |\psi^{\pm}\rangle^{F} |\psi^{\pm}\rangle^{S}$. } \label{fig10}
\end{figure}

\begin{figure}[th]%[tpb]
\centering
\includegraphics[width=6cm,angle=0]{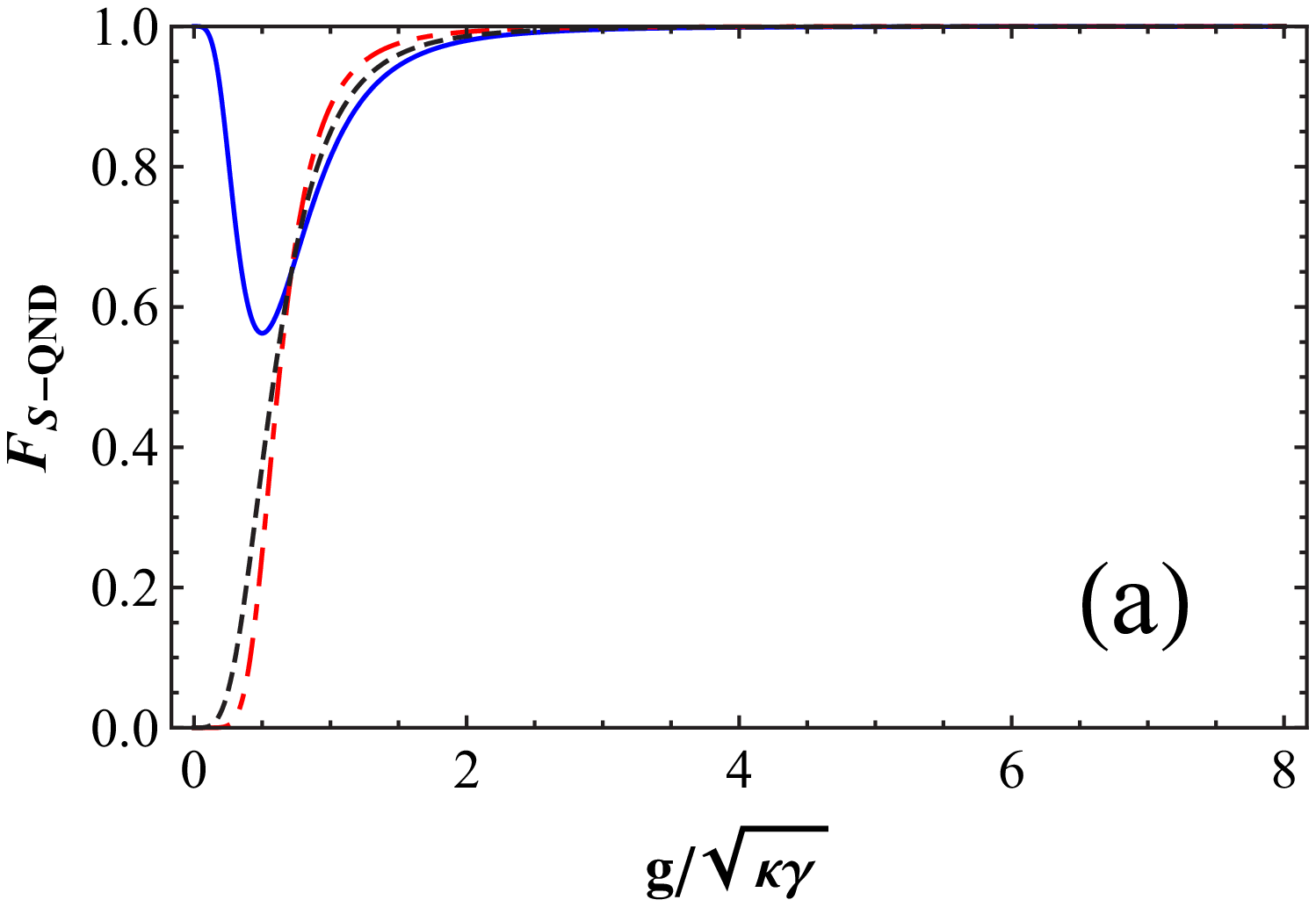} \hspace{30pt}
\includegraphics[width=6cm,angle=0]{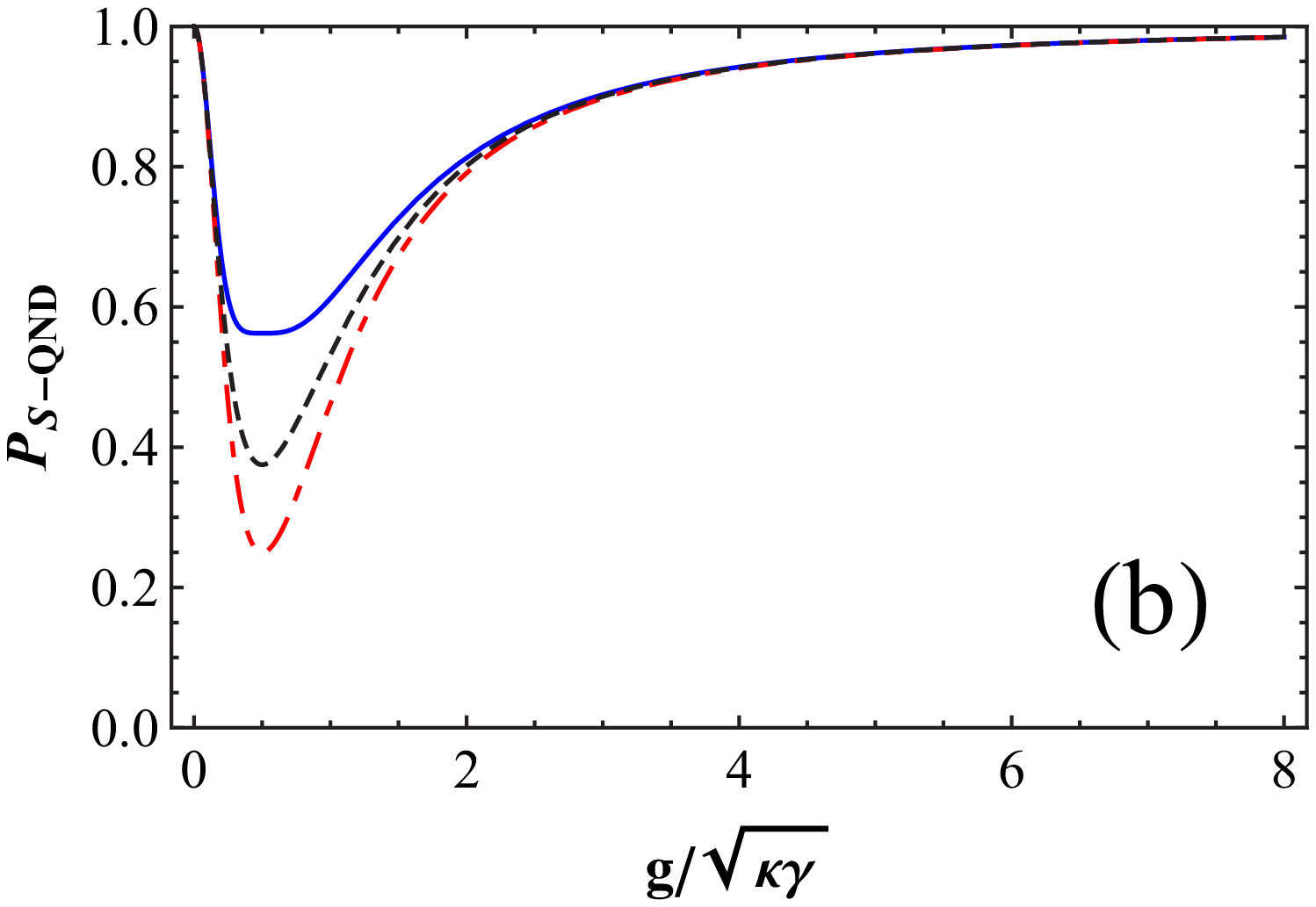}
\caption{(a) The fidelity of the spatial-mode-parity check QND. (b) The efficiency of the spatial-mode-parity check QND.
The solid line represents $F_{1}(\eta_{1})$ for $|\phi^{\pm}\rangle^{p} |\phi^{\pm}\rangle^{F} |\phi^{\pm}\rangle^{F}$.
The dash-dotted line represents $F_{S2}(\eta_{S2})=F_{S4}(\eta_{S4})=F_{S6}(\eta_{S6})=F_{S7}(\eta_{S7})=F_{S8}(\eta_{S8})$
for $|\psi^{\pm}\rangle^{p} |\phi^{\pm}\rangle^{F} |\phi^{\pm}\rangle^{S}$, $|\psi^{\pm}\rangle^{p} |\phi^{\pm}\rangle^{F} |\psi^{\pm}\rangle^{S}$,
$|\psi^{\pm}\rangle^{p} |\psi^{\pm}\rangle^{F} |\phi^{\pm}\rangle^{S}$, $|\phi^{\pm}\rangle^{p} |\psi^{\pm}\rangle^{F} |\psi^{\pm}\rangle^{S}$,
and $|\psi^{\pm}\rangle^{p} |\psi^{\pm}\rangle^{F} |\psi^{\pm}\rangle^{S}$. The dashed line represents $F_{3}(\eta_{3})=F_{5}(\eta_{5})$
for $|\phi^{\pm}\rangle^{p} |\phi^{\pm}\rangle^{F} |\psi^{\pm}\rangle^{S}$
and $|\phi^{\pm}\rangle^{p} |\psi^{\pm}\rangle^{F} |\phi^{\pm}\rangle^{S}$.} \label{fig11}
\end{figure}

\begin{figure}[th]%[tpb]
\centering
\includegraphics[width=6cm,angle=0]{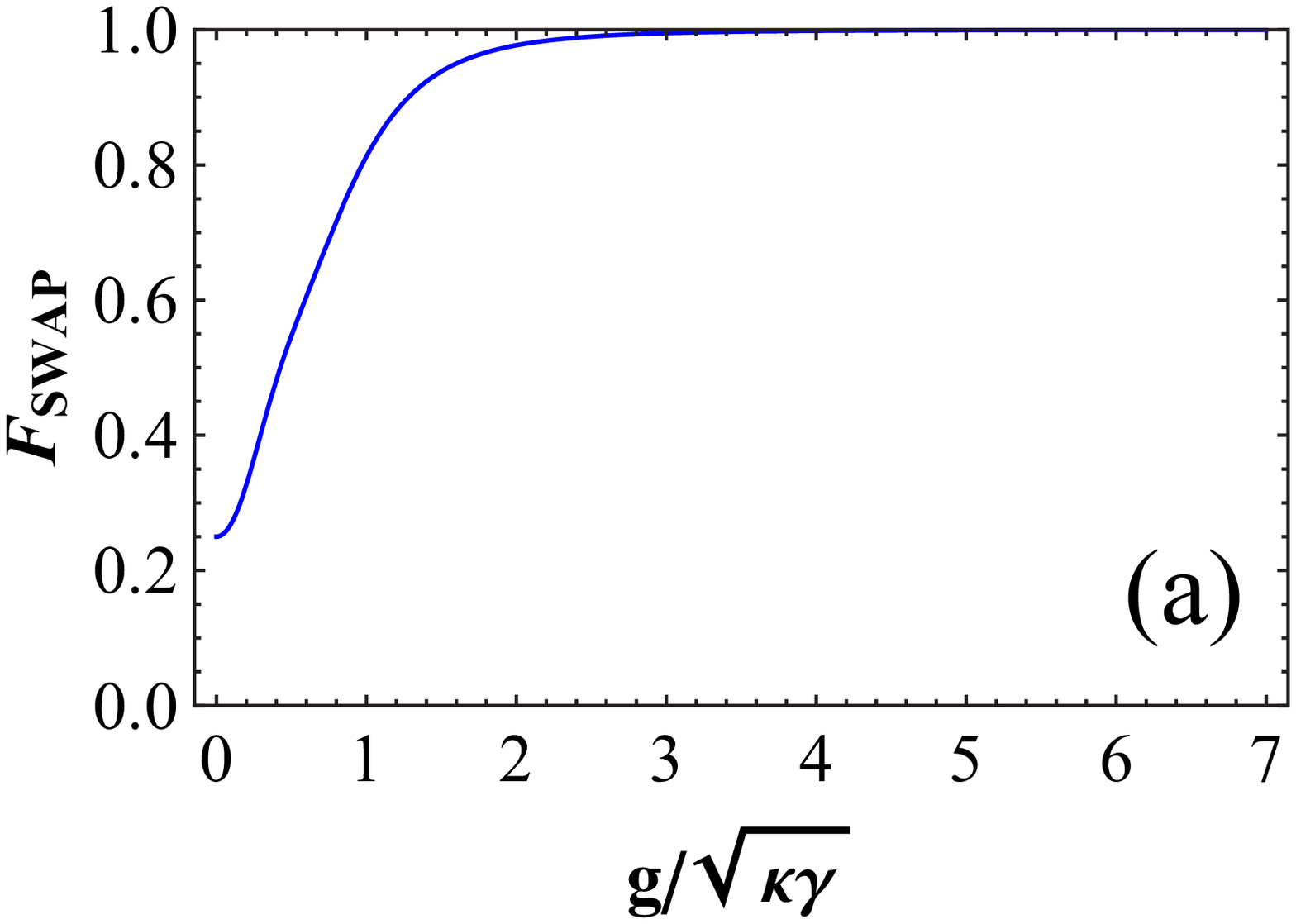} \hspace{30pt}
\includegraphics[width=6cm,angle=0]{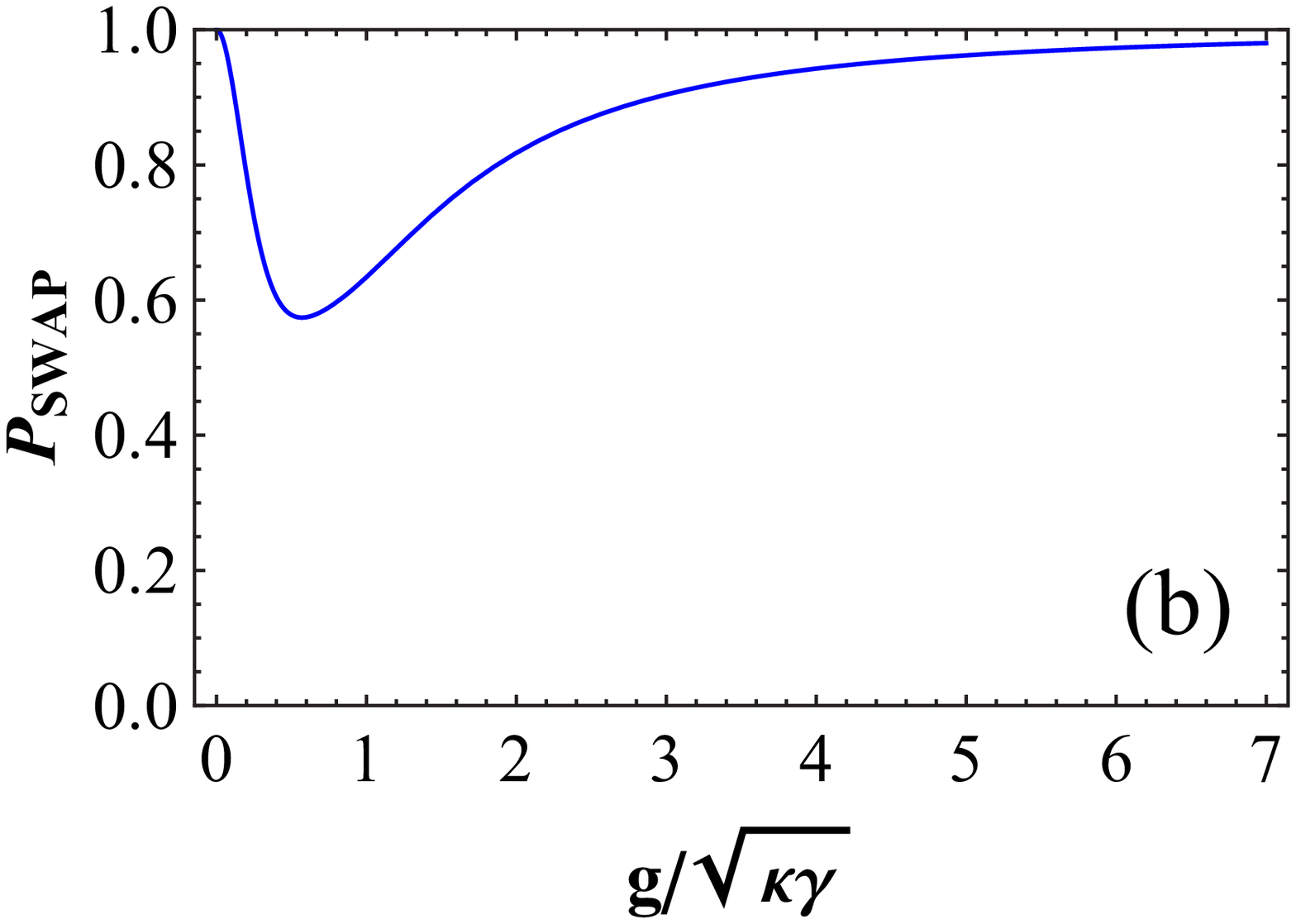}
\caption{(a) The fidelity of the P-P-SWAP gate. (b) The efficiency of the P-P-SWAP gate. } \label{fig12}
\end{figure}

\section{Discussion and summary}
\label{sec7}

With the P-S-QND and the SWAP gates, we construct our efficient
hyper-EPP for the mixed hyperentangled Bell states in the
polarization DOF and two longitudinal-momentum DOFs with bit-flip
errors. For constructing the P-S-QND and the SWAP gates, we utilize
the reflection coefficient produced by the nonlinear interaction
between the single photon and the NV-cavity system. Under the
resonant condition $\omega_{0}=\omega_{p}=\omega_{c}$, the
reflection coefficient of the input photon is affected by the
coupling strength $g$, the NV decay rate $\gamma$, and the cavity
damping rate $\kappa$. If we neglect $\kappa$ and $\gamma$, the
fidelities and the efficiencies of the P-S-QND and the SWAP gates
can reach $100\%$. However, in the realistic condition, the
fidelities and the efficiencies are suppressed by $\gamma$ and
$\kappa$. The fidelity is defined as $F=|\langle \varphi_{r} |
\varphi_{i} \rangle|^{2}$. Here $| \varphi_{i} \rangle$ represents
the final state in the ideal condition and $|\varphi_{r}\rangle$
represents the final state in the realistic condition. The
efficiency is defined as $\eta=n_{out}/n_{in}$, where $n_{out}$
represents the number of the output photons and $n_{in}$ represents
the number of the input photons. For the polarization parity-check
QND, there exist two fidelities $F_{Pi}$ $(i=1,2)$ and two
efficiencies $\eta_{pi}$ $(i=1,2)$ corresponding to the two
different polarization states $|\phi^{\pm}\rangle^{p}$ and
$|\psi^{\pm}\rangle^{p}$. For the spatial-mode parity-check QND,
there exist eight fidelities $F_{Si}$ $(i=1\sim8)$ and eight
efficiencies $\eta_{Si}$ $(i=1\sim8)$ corresponding to the states
$|\phi^{\pm}\rangle^{p} |\phi^{\pm}\rangle^{F}
|\phi^{\pm}\rangle^{S}$, $|\psi^{\pm}\rangle^{p}
|\phi^{\pm}\rangle^{F} |\phi^{\pm}\rangle^{S}$,
$|\phi^{\pm}\rangle^{p} |\phi^{\pm}\rangle^{F}
|\psi^{\pm}\rangle^{S}$, $|\psi^{\pm}\rangle^{p}
|\phi^{\pm}\rangle^{F} |\psi^{\pm}\rangle^{S}$,
$|\phi^{\pm}\rangle^{p} |\psi^{\pm}\rangle^{F}
|\phi^{\pm}\rangle^{S}$, $|\psi^{\pm}\rangle^{p}
|\psi^{\pm}\rangle^{F} |\phi^{\pm}\rangle^{S}$,
$|\phi^{\pm}\rangle^{p} |\psi^{\pm}\rangle^{F}
|\psi^{\pm}\rangle^{S}$, and $|\psi^{\pm}\rangle^{p}
|\psi^{\pm}\rangle^{F} |\psi^{\pm}\rangle^{S}$, respectively. As
$r_{0}$ is not affected by $g$, $\kappa$, nor $\gamma$, we have
$F_{S2}(\eta_{S2})=F_{S4}(\eta_{S4})=F_{S6}(\eta_{S6})=F_{S7}(\eta_{S7})=F_{S8}(\eta_{S8})$
and $F_{S3}(\eta_{S3})=F_{S5}(\eta_{S5})$. The fidelities and the
efficiencies (discussed the detail in the Appendix) of the polarization
parity-check QND, spatial parity-check QND, and the P-P-SWAP gate,
which vary with the parameter $g/\sqrt{\kappa\gamma}$, are shown in
Fig.~\ref{fig10}, Figs.~\ref{fig11}, and  ~\ref{fig12},
respectively.

By far, several groups have experimentally demonstrated the coupling
between a diamond NV center and a microcavity
\cite{NV8,NV9,NV10,NV11}. In 2009, Barclay \emph{et al.} \cite{NV8}
reported their experiment with the parameters
$[g,\kappa,\gamma_{tot},\gamma]/2\pi=[0.30,26,0.013,0.0004]GHz$ for
coupling the diamond NV centers to a chip-based microcavity. This
experiment is implemented in a weak coupling regime with a low-$Q$
factor. In their experiment, the NV center coherent coupling rate within
the narrow-band zero phonon line (ZPL) is almost one-third of the
total coupling rate. We have $r=0.94$ at
$\omega_{0}=\omega_{p}=\omega_{c}$. Based on their experimental
parameters, the fidelities and the efficiencies of our polarization
parity-check QND can reach $F_{p1,2}=99.76\%,99.91\%$ and
$\eta_{p1,2}=94.84\%,94.54\%$, respectively. The fidelities and the efficiencies
of our spatial-mode parity-check QND can reach
$F_{S1,2,3}=99.53\%,99.83\%,99.68\%$ and
$\eta_{S1,2,3}=89.95\%,89.38\%,89.66\%$, respectively. The fidelity and the
efficiency of the P-P-SWAP gate can reach $F_{SWAP}=99.46\%$ and
$\eta_{SWAP}=90.08\%$, respectively.

In our proposal, the two cavity modes with right- and left- circular
polarizations, which couple to the two transitions
$|+\rangle\leftrightarrow|A_{2}\rangle$ and
$|-\rangle\leftrightarrow|A_{2}\rangle$ respectively, are required.
Many good experiments that provide a cavity supporting both of two
circularly-polarized modes with the same frequency have been
realized \cite{modes1,modes2,modes3,modes4,modes5,modes6,modes7}.
For example, Luxmoore \emph{et al.} \cite{modes1} presented a
technique for fine tuning of the energy split between the two
circularly-polarized modes to just 0.15nm in 2012.

In summary, we have proposed a hyper-EPP for the mixed
hyperentangled Bell states with bit-flip errors in three DOFs,
including the polarization and two longitudinal-momentum DOFs.  Our
hyper-EPP is completed with one or two steps. In the first step,
Alice and Bob perform the P-S-QNDs and Hadamard operations on their
own photons. If the states of the two photon pairs are projected
into the states in case (1), the hyperentanglement purification
process is accomplished and it costs  one photon pair to purify
another one. If the states of the two photon pairs are projected
into   cases (3), (4), (5), (6), (7) and (8), the second step of
our hyper-EPP is needed. In the second step, Alice and Bob swap the
states between two-photon systems in the same DOF or between
different DOFs of the one photon by the method of combining the
P-P-SWAP gate, the P-F-SWAP gate, and the P-S-SWAP gate. If the
hyperentanglement purification process is accomplished with two
steps, it costs another three  pairs to purify one photon pair. We
have showed the feasibility of our hyper-EPP as high fidelities and
efficiencies of P-S-QND and SWAP gates can be obtained based on the
existing experimental parameters and there exists the cavity that
supports two circularly-polarized modes with the same frequency.

\section*{ACKNOWLEDGEMENTS}

This work was supported by the National Natural Science Foundation
of China under Grant No. 11674033 and No. 11474026, and the
Fundamental Research Funds for the Central Universities under Grant
No. 2015KJJCA01.

\section*{APPENDIX}
\label{sec8}

\subsection{Case (2) in the first step of our  hyper-EPP with SWAP gates}

In case (2), the two identical two-photon systems $AC$ and $BD$ are
in different parity-mode in all of the three DOFs, including the
polarization DOF, the first-longitudinal-momentum DOF, and the second-longitudinal- momentum DOF.

When the photons $AC$ are in the even-parity modes and $BD$ are in
the odd-parity modes at the same time in the
polarization DOF, in the first-longitudinal-momentum DOF, and in the
second-longitudinal-momentum DOF, Alice and Bob obtain the states
\begin{equation}\begin{split}
&|\Phi_{5}\rangle^{p}=\frac{1}{\sqrt{2}}(|RRRL\rangle+|LLLR\rangle)_{ABCD},\\
&|\Phi_{6}\rangle^{p}=\frac{1}{\sqrt{2}}(|RLRR\rangle+|LRLL\rangle)_{ABCD},\\
&|\Phi_{5}\rangle^{F}=\frac{1}{\sqrt{2}}(|rrrl\rangle+|lllr\rangle)_{ABCD},\\
&|\Phi_{6}\rangle^{F}=\frac{1}{\sqrt{2}}(|rlrr\rangle+|lrll\rangle)_{ABCD},\\
&|\Phi_{5}\rangle^{S}=\frac{1}{\sqrt{2}}(|EEEI\rangle+|IIIE\rangle)_{ABCD},\\
&|\Phi_{6}\rangle^{S}=\frac{1}{\sqrt{2}}(|EIEE\rangle+|IEII\rangle)_{ABCD}.
\end{split}\end{equation}

On the contrary, when the photons $AC$ are in the odd-parity modes
and $BD$ are in the even-parity modes at the same time in the
polarization DOF, in the first-longitudinal-momentum DOF
or in the second-longitudinal-momentum DOF, Alice and Bob obtain the states
\begin{equation}\begin{split}
&|\Phi_{7}\rangle^{p}=\frac{1}{\sqrt{2}}(|RRLR\rangle+|LLRL\rangle)_{ABCD},\\
&|\Phi_{8}\rangle^{p}=\frac{1}{\sqrt{2}}(|RLLL\rangle+|LRRR\rangle)_{ABCD},\\
&|\Phi_{7}\rangle^{F}=\frac{1}{\sqrt{2}}(|rrlr\rangle+|llrl\rangle)_{ABCD},\\
&|\Phi_{8}\rangle^{F}=\frac{1}{\sqrt{2}}(|rlll\rangle+|lrrr\rangle)_{ABCD},\\
&|\Phi_{7}\rangle^{S}=\frac{1}{\sqrt{2}}(|EEIE\rangle+|IIEI\rangle)_{ABCD},\\
&|\Phi_{8}\rangle^{S}=\frac{1}{\sqrt{2}}(|EIII\rangle+|IEEE\rangle)_{ABCD}.
\end{split}\end{equation}
The states $|\Phi_{7}\rangle^{p}$, $|\Phi_{8}\rangle^{p}$,
$|\Phi_{7}\rangle^{F}$, $|\Phi_{8}\rangle^{F}$,
$|\Phi_{7}\rangle^{S}$, and $|\Phi_{8}\rangle^{S}$ can be
transformed into states $|\Phi_{5}\rangle^{p}$,
$|\Phi_{6}\rangle^{p}$, $|\Phi_{5}\rangle^{F}$,
$|\Phi_{6}\rangle^{F}$, $|\Phi_{5}\rangle^{S}$, and
$|\Phi_{6}\rangle^{S}$, respectively.

Next, Alice and Bob perform Hadamard operations in the polarization
and the two longitudinal-momentum DOFs on photons $C$ and $D$,
respectively. In the polarization DOF, the states
$|\Phi_{5}\rangle^{p}$ and $|\Phi_{6}\rangle^{p}$ are transformed
into the states $|\Phi_{5'}\rangle^{p}$ and $|\Phi_{6'}\rangle^{p}$,
respectively. In the first-longitudinal-momentum DOF, the states
$|\Phi_{5}\rangle^{F}$ and $|\Phi_{6}\rangle^{F}$ are transformed
into $|\Phi_{5}'\rangle^{F}$ and $|\Phi_{6}'\rangle^{F}$,
respectively. In the second-longitudinal-momentum DOF, the states
$|\Phi_{5}\rangle^{S}$ and $|\Phi_{6}\rangle^{S}$ are transformed
into $|\Phi_{5}'\rangle^{S}$ and $|\Phi_{6}'\rangle^{S}$,
respectively. Here
\begin{equation}\begin{split}
|\Phi'_{5}\rangle^{p}\;=\;&\frac{1}{2\sqrt{2}}[(|RR\rangle+|LL\rangle)_{AB}(|RR\rangle-|LL\rangle)_{CD}\\
&+(|RR\rangle-|LL\rangle)_{AB}(-|RL\rangle+|LR\rangle)_{CD}],\\
|\Phi'_{6}\rangle^{p}\;=\;&\frac{1}{2\sqrt{2}}[(|RL\rangle+|LR\rangle)_{AB}(|RR\rangle+|LL\rangle)_{CD}\\
&+(-|RL\rangle+|LR\rangle)_{AB}(-|RL\rangle-|LR\rangle)_{CD}],\\
|\Phi'_{5}\rangle^{F}\;=\;&\frac{1}{2\sqrt{2}}[(|rr\rangle+|ll\rangle)_{AB}(|rr\rangle-|ll\rangle)_{CD}\\
&+(|rr\rangle-|ll\rangle)_{AB}(-|rl\rangle+|lr\rangle)_{CD}],\\
|\Phi'_{6}\rangle^{F}\;=\;&\frac{1}{2\sqrt{2}}[(|rl\rangle+|lr\rangle)_{AB}(|rr\rangle+|ll\rangle)_{CD}\\
&+(-|rl\rangle+|lr\rangle)_{AB}(-|rl\rangle-|lr\rangle)_{CD}],\\
|\Phi'_{5}\rangle^{S}\;=\;&\frac{1}{2\sqrt{2}}[(|EE\rangle+|II\rangle)_{AB}(|EE\rangle-|II\rangle)_{CD}\\
&+(|EE\rangle-|II\rangle)_{AB}(-|EI\rangle+|IE\rangle)_{CD}],\\
|\Phi'_{6}\rangle^{S}\;=\;&\frac{1}{2\sqrt{2}}[(|EI\rangle+|IE\rangle)_{AB}(|EE\rangle+|II\rangle)_{CD}\\
&+(-|EI\rangle+|IE\rangle)_{AB}(-|EI\rangle-|IE\rangle)_{CD}].
\end{split}\end{equation}

Finally, the photons $C$ and $D$ are detected by
single-photon detectors, respectively. If the photons $CD$ are in the even-parity
mode in the polarization DOF (the first-longitudinal-momentum DOF or the
second-longitudinal-momentum DOF), nothing is needed to be done. If the
photons $CD$ are in the odd-parity mode in the polarization DOF (the
first-longitudinal-momentum DOF or the second-longitudinal-momentum
DOF), the phase-flip operation $\sigma_{z}^{p}=|R\rangle\langle
R|-|L\rangle \langle L|$ ($\sigma_{z}^{F}=|r\rangle\langle
r|-|l\rangle \langle l|$ or $\sigma_{z}^{S}=|E\rangle\langle
E|-|I\rangle \langle I|$) is operated on the photon $B$. If  $AC$ and $BD$ are projected into this case, Alice and Bob will discard the two photon pairs.

\subsection{The complete second step of our  hyper-EPP with SWAP gates}

(a) If the two two-photon systems $AB$ and $CD$ are projected into
the states in case (3) in the first step, Alice and Bob can seek
another two two-photon systems $A'B'$ and $C'D'$ which are projected
into the states in case (8). The P-P-SWAP gate shown in
Fig.~\ref{fig5} can be used to complete the second step of our
hyper-EPP. The states of the two photon pairs $AB$ and $A'B'$ are
transformed into the states
\begin{equation}\begin{split}
&|\Phi\rangle_{AB5}=|\phi^{+}\rangle^{p}|\phi^{+}\rangle^{F}|\phi^{+}\rangle^{S}, \\
&|\Phi\rangle_{A'B'5}=|\psi^{+}\rangle^{p}|\psi^{+}\rangle^{F}|\psi^{+}\rangle^{S}.
\end{split}\end{equation}
Here $|\Phi\rangle_{AB5}$ and $|\Phi\rangle_{A'B'5}$ are the finale states of case (a) in the second step of our hyper-EPP.

(b) If the two two-photon systems $AB$ and $CD$ are projected into
the states in case (4) in the first step, Alice and Bob can seek
another two two-photon systems $A'B'$ and $C'D'$ which are projected
into the states in case (7). The P-P-SWAP gate shown in
Fig.~\ref{fig5} and the P-F-SWAP gate shown in Fig.~\ref{fig6}(a)
can be used to complete the second step of our hyper-EPP.

First, Alice swaps the polarization states and the first-longitudinal-momentum states of photons $A$ and performs the same
operations on photon $A'$. Bob performs the same operations on
photons $B$ and $B'$, respectively. Second, Alice swaps the
polarization states between the photons $A$ and $A'$. Bob swaps the
polarization states between the photons $B$ and $B'$. Finally, Alice
and Bob swap the polarization states and the first-longitudinal-momentum states on the photons $A$, $B$, $A'$ and $B'$ again. After
these operations, the states of photon pairs become
\begin{equation}\begin{split}
&|\Phi\rangle_{AB6}=|\psi^{+}\rangle^{p}|\psi^{+}\rangle^{F}|\phi^{+}\rangle^{S},\\
&|\Phi\rangle_{A'B'6}=|\phi^{+}\rangle^{p}|\phi^{+}\rangle^{F}|\psi^{+}\rangle^{S}.
\end{split}\end{equation}
Here $|\Phi\rangle_{AB6}$ and $|\Phi\rangle_{A'B'6}$ are the final states of case (b) in the second step of our hyper-EPP.

(c) If the two two-photon systems $AB$ and $CD$ are projected into
the states in case (5) in the first step, Alice and Bob can seek
another two two-photon systems $A'B'$ and $C'D'$ which are projected
into the states in case (6). Alice and Bob perform the similar
operations as what they do in case (b) in the second step. The
difference is that in this case, they use the P-S-SWAP gate shown in
Fig.~\ref{fig6}(b), instead of the P-F-SWAP gate. The final states of
case (c) in the second step of our hyper-EPP are
\begin{equation}\begin{split}
&|\Phi\rangle_{AB7}=|\psi^{+}\rangle^{p}|\phi^{+}\rangle^{F}|\psi^{+}\rangle^{S},\\
&|\Phi\rangle_{A'B'7}=|\phi^{+}\rangle^{p}|\psi^{+}\rangle^{F}|\phi^{+}\rangle^{S}.
\end{split}\end{equation}

(d) If the two two-photon systems $AB$ and $CD$ are projected into
the states in case (7) in the first step, Alice and Bob can seek
another two two-photon systems $A'B'$ and $C'D'$ which are projected
into the states in case (4). The second step of our hyper-EPP is
completed with the P-P-SWAP gates and the P-S-SWAP gates.

First, Alice and Bob swap the polarization states of the two
two-photon system $AB$ and the states of the system $A'B'$.  Second,
Alice and Bob swap the polarization states and the second
longitudinal-momentum states of photons $A$, $B$, $A'$, and $B'$,
respectively. Third, Alice and Bob swap the polarization states of
the two systems $AB$ and $A'B'$ again. Finally, Alice and Bob swap
the polarization states of the second-longitudinal-momentum states
of photons $A$, $B$, $A'$, and $B'$, respectively. The states of the
two systems are changed to be
\begin{equation}\begin{split}
&|\Phi\rangle_{AB8}=|\phi^{+}\rangle^{p}|\phi^{+}\rangle^{F}|\psi^{+}\rangle^{S}\\
&|\Phi\rangle_{A'B'8}=|\psi^{+}\rangle^{p}|\psi^{+}\rangle^{F}|\phi^{+}\rangle^{S}.
\end{split}\end{equation}
Here $|\Phi\rangle_{AB8}$ and $|\Phi\rangle_{A'B'8}$ are the final states of this case in the second step of our hyper-EPP.

(e) If the two two-photon systems $AB$ and $CD$ are projected into
the states in case (8) in the first step, Alice and Bob can seek
another two two-photon systems $A'B'$ and $C'D'$ which are projected
into the states in case (3). The second step of our hyper-EPP is
completed with the P-P-SWAP gates, P-F-SWAP gates, and P-S-SWAP
gates.

First, Alice and Bob swap the polarization states and the first-longitudinal-momentum states of photons $A$, $B$, $A'$, and $B'$ by the
P-F-SWAP gate, respectively. Second, Alice and Bob swap the
polarization states between system $AB$ and system $A'B'$  with the
P-P-SWAP gate. Third, Alice and Bob swap the polarization states and
the first-longitudinal-momentum states of photons $A$, $B$, $A'$, and
$B'$ again with the P-F-SWAP gate, respectively. Fourth, Alice and Bob
swap the polarization states and the second-longitudinal-momentum
states of photons $A$, $B$, $A'$, and $B'$ with the P-S-SWAP gate,
respectively. Fifth, Alice and Bob swap the polarization states
between system $AB$ and system $A'B'$ by the P-P-SWAP gate for the third
time. Finally, Alice and Bob swap the polarization states and the
second-longitudinal-momentum states of photons $A$, $B$, $A'$, and
$B'$ respectively again. Thus, the states of systems become
\begin{equation}\begin{split}
&|\Phi\rangle_{AB9}=|\psi^{+}\rangle^{p}|\psi^{+}\rangle^{F}|\psi^{+}\rangle^{S}\\
&|\Phi\rangle_{A'B'9}=|\phi^{+}\rangle^{p}|\phi^{+}\rangle^{F}|\phi^{+}\rangle^{S}.
\end{split}\end{equation}
Here $|\Phi\rangle_{AB9}$ and $|\Phi\rangle_{A'B'9}$ are the final states of this case in the second step of our hyper-EPP.

\begin{widetext}

\subsection{The fidelity and the efficiency}

There exist two fidelities $F_{P1,2}$ and efficiencies $\eta_{P1,2}$
of the polarization parity-check QND. The fidelity (efficiency)
$F_{P1}(\eta_{P1})$ corresponds to the states
$|\phi^{\pm}\rangle^{p} |\phi^{\pm}\rangle^{F}
|\phi^{\pm}\rangle^{S}$, $|\phi^{\pm}\rangle^{p}
|\phi^{\pm}\rangle^{F} |\psi^{\pm}\rangle^{S}$,
$|\phi^{\pm}\rangle^{p} |\psi^{\pm}\rangle^{F}
|\phi^{\pm}\rangle^{S}$, and $|\phi^{\pm}\rangle^{p}
|\psi^{\pm}\rangle^{F} |\psi^{\pm}\rangle^{S}$. The fidelity
(efficiency) $F_{P2}(\eta_{P2})$ corresponds to the states
$|\psi^{\pm}\rangle^{p} |\phi^{\pm}\rangle^{F}
|\phi^{\pm}\rangle^{S}$, $|\psi^{\pm}\rangle^{p}
|\phi^{\pm}\rangle^{F} |\psi^{\pm}\rangle^{S}$,
$|\psi^{\pm}\rangle^{p} |\psi^{\pm}\rangle^{F}
|\phi^{\pm}\rangle^{S}$, and $|\psi^{\pm}\rangle^{p}
|\psi^{\pm}\rangle^{F} |\psi^{\pm}\rangle^{S}$. The fidelities of
the polarization parity-check QND can be expressed as
\begin{equation}\begin{split}
&F_{P1}=\frac{|2+r^{2}+r_{0}^{2}|^{2}}{4[2+|r^{2}|^{2}+|r_{0}^{2}|^{2}]},\;\;\;\;\;\;\;\;\;\;
F_{P2}=\frac{|r-r_{0}|^{2}}{2[|r|^{2}+|r_{0}|^{2}]}.
\end{split}\end{equation}
The efficiencies of the polarization parity-check QND can be expressed as
\begin{equation}\begin{split}
&\eta_{p1}=\frac{1}{4}[2+|r^{2}|^{2}+|r_{0}^{2}|^{2}],
\;\;\;\;\;\;\;\;\;\; \eta_{P2}=\frac{1}{2}[|r_{0}|^{2}+|r|^{2}].
\end{split}\end{equation}
The fidelity and the efficiency of the polarization parity-check
QND, which vary with the parameter $g/\sqrt{\kappa\gamma}$, are
shown in Figs.~\ref{fig9}(a) and  \ref{fig9}(b), respectively.

For the spatial-mode parity-check QND, there exist eight fidelities
$F_{Si}(i=1,2,\cdots, 8)$ and eight efficiencies
$\eta_{Si}(i=1,2,\cdots, 8)$ corresponding to the states
$|\phi^{\pm}\rangle^{p} |\phi^{\pm}\rangle^{F}
|\phi^{\pm}\rangle^{F}$, $|\psi^{\pm}\rangle^{p}
|\phi^{\pm}\rangle^{F} |\phi^{\pm}\rangle^{F}$,
$|\phi^{\pm}\rangle^{p} |\phi^{\pm}\rangle^{F}
|\psi^{\pm}\rangle^{F}$, $|\psi^{\pm}\rangle^{p}
|\phi^{\pm}\rangle^{F} |\psi^{\pm}\rangle^{F}$,
$|\phi^{\pm}\rangle^{p} |\psi^{\pm}\rangle^{F}
|\phi^{\pm}\rangle^{F}$, $|\psi^{\pm}\rangle^{p}
|\psi^{\pm}\rangle^{F} |\phi^{\pm}\rangle^{F}$,
$|\phi^{\pm}\rangle^{p} |\psi^{\pm}\rangle^{F}
|\psi^{\pm}\rangle^{F}$, and $|\psi^{\pm}\rangle^{p}
|\psi^{\pm}\rangle^{F} |\psi^{\pm}\rangle^{F}$, respectively. The
fidelities of the spatial-mode parity-check QND can be expressed as
\begin{equation}\begin{split}
&F_{S1}=\frac{|(r^{2}+r_{0}^{2})^{2}+4(r^{2}+r_{0}^{2})+4|^{2}}{16[4(|r^{2}|^{2}+|r_{0}^{2}|^{2})+(|r^{4}|^{2}+2|r^{2}r_{0}^{2}|^{2}+|r_{0}^{4}|^{2})+4]},\\
&F_{S2}=\frac{|r^{2}r_{0}^{2}-2rr_{0}+1|^{2}}{4[|r^{2}r_{0}^{2}|^{2}+2|rr_{0}|^{2}+1]},\\
&F_{S3}=\frac{|(r^{2}+r_{0}^{2})(r-r_{0})+2(r-r_{0})|^{2}}{8[(|r^{3}|^{2}+|r_{0}^{3}|^{2}+|r_{0}r^{2}|^{2}+|r_{0}^{2}r|^{2})+2(|r|^{2}+|r_{0}|^{2})]},\\
&F_{S4}=\frac{|(r-r_{0})(1-rr_{0})|^{2}}{4[|rr_{0}^{2}|^{2}+|r_{0}r^{2}|^{2}+|r|^{2}+|r_{0}|^{2}]},\\
&F_{S5}=\frac{|(r^{2}+r_{0}^{2})(r-r_{0})+2(r-r_{0})|^{2}}{8[(|r^{3}|^{2}+|rr_{0}^{2}|^{2}+|r_{0}r^{2}|^{2}+|r_{0}^{3}|^{2})+2(|r|^{2}+|r_{0}|^{2})]},\\
&F_{S6}=\frac{|(r-r_{0})(1-rr_{0})|^{2}}{4[|r|^{2}+|r_{0}|^{2}+|rr_{0}^{2}|^{2}+|r_{0}r^{2}|^{2}]},\\
&F_{S7}=\frac{|r^{2}+r_{0}^{2}-2rr_{0}|^{2}}{4[|r^{2}|^{2}+2|rr_{0}|^{2}+|r_{0}^{2}|^{2}]},\\
&F_{S8}=\frac{|r^{2}+r_{0}^{2}-2rr_{0}|^{2}}{4[|r^{2}|^{2}+2|rr_{0}|^{2}+|r_{0}^{2}|^{2}]}.
\end{split}\end{equation}
The efficiencies of the spatial-mode parity-check QND can be expressed
as
\begin{equation}\begin{split}
&\eta_{S1}=\frac{1}{16}[4(|r^{2}|^{2}+|r_{0}^{2}|^{2})+(|r^{4}|^{2}+2|r^{2}r_{0}^{2}|^{2}+|r_{0}^{4}|^{2})+4],\\
&\eta_{S2}=\frac{1}{4}[|r^{2}r_{0}^{2}|^{2}+2|rr_{0}|^{2}+1],\\
&\eta_{S3}=\frac{1}{8}[(|r^{3}|^{2}+|r_{0}^{3}|^{2}+|r_{0}r^{2}|^{2}+|r_{0}^{2}r|^{2})+2(|r|^{2}+|r_{0}|^{2})],\\
&\eta_{S4}=\frac{1}{4}[|rr_{0}^{2}|^{2}+|r_{0}r^{2}|^{2}+|r|^{2}+|r_{0}|^{2}],\\
&\eta_{S5}=\frac{1}{8}[(|r^{3}|^{2}+|rr_{0}^{2}|^{2}+|r_{0}r^{2}|^{2}+|r_{0}^{3}|^{2})+2(|r|^{2}+|r_{0}|^{2})],\\
&\eta_{S6}=\frac{1}{4}[|r|^{2}+|r_{0}|^{2}+|rr_{0}^{2}|^{2}+|r_{0}r^{2}|^{2}],\\
&\eta_{S7}=\frac{1}{4}[|r^{2}|^{2}+2|rr_{0}|^{2}+|r_{0}^{2}|^{2}],\\
&\eta_{S8}=\frac{1}{4}[|r^{2}|^{2}+2|rr_{0}|^{2}+|r_{0}^{2}|^{2}].
\end{split}\end{equation}

As $r_{0}=-1$ is an exact value, we have
$F_{S2}(\eta_{S2})=F_{S4}(\eta_{S4})=F_{S6}(\eta_{S6})=F_{S7}(\eta_{S7})=F_{S8}(\eta_{S8})$
and $F_{S3}(\eta_{S3})=F_{S5}(\eta_{S5})$. The fidelities and the
efficiencies of the spatial-mode parity-check QND, which vary with
the parameter $g/\sqrt{\kappa\gamma}$, are shown in
Figs.~\ref{fig10}(a) and  \ref{fig10}(b), respectively.

The fidelity and the efficiency of the P-P-SWAP gate proposed in
Sec.IV(A) can be expressed as
\begin{equation}\begin{split}
F'_{SWAP}=&\frac{|f'_{1}+f'_{2}+f'_{3}+f'_{4}+f'_{4}+f'_{5}+f'_{6}+f'_{7}+f'_{8}|^{2}}{\eta'_{SWAP}},\\
\eta'_{SWAP}=&\frac{1}{32}[|2\alpha_{1}\alpha_{2}r^{2}+(\alpha_{1}\beta_{2}+\beta_{1}\alpha_{2})(r^{2}r_{0}+r^{3}+r_{0}^{3}-r_{0}^{2}r)+\beta_{1}\beta_{2}(r^{4}+r_{0}^{4})|^{2}\\
&+|2\alpha_{1}\alpha_{2}r^{2}+(\alpha_{1}\beta_{2}+\beta_{1}\alpha_{2})(r^{2}r_{0}+r^{3}-r_{0}^{3}+r_{0}^{2}r)+\beta_{1}\beta_{2}(r^{4}-r_{0}^{4}+2r^{2}r_{0}^{2})|^{2}\\
&+|2\alpha_{1}\alpha_{2}r-(\alpha_{1}\beta_{2}-\beta_{1}\alpha_{2})(r^{2}+r_{0}^{2})-\beta_{1}\beta_{2}(rr_{0}^{2}+r^{3}+r_{0}^{3}-r_{0}r^{2})|^{2}\\
&+|2\alpha_{1}\alpha_{2}r-(\alpha_{1}\beta_{2}-\beta_{1}\alpha_{2})(2rr_{0}+r^{2}-r_{0}^{2})-\beta_{1}\beta_{2}(rr_{0}^{2}+r^{3}-r_{0}^{3}+r_{0}r^{2})|^{2}\\
&+|2\alpha_{1}\alpha_{2}r+(\alpha_{1}\beta_{2}-\beta_{1}\alpha_{2})(r^{2}+r_{0}^{2})-\beta_{1}\beta_{2}(rr_{0}^{2}+r^{3}+r_{0}^{3}-r_{0}r^{2})|^{2}\\
&+|2\alpha_{1}\alpha_{2}r+(\alpha_{1}\beta_{2}-\beta_{1}\alpha_{2})(2rr_{0}+r^{2}-r_{0}^{2})-\beta_{1}\beta_{2}(rr_{0}^{2}+r^{3}-r_{0}^{3}+r_{0}r^{2})|^{2}\\
&+|2\alpha_{1}\alpha_{2}-2(\alpha_{1}\beta_{2}+\beta_{1}\alpha_{2})r_{0}+2\beta_{1}\beta_{2}r_{0}^{2}|^{2}\\
&+|2\alpha_{1}\alpha_{2}-2(\alpha_{1}\beta_{2}+\beta_{1}\alpha_{2})r+2\beta_{1}\beta_{2}r^{2}|^{2}].
\end{split}\end{equation}
%\end{widetext}
Here,
\begin{equation}\begin{split}
&f'_{1}=\frac{1}{16}[2\alpha_{1}\alpha_{2}r^{2}+(\alpha_{1}\beta_{2}+\beta_{1}\alpha_{2})(r^{2}r_{0}+r^{3}+r_{0}^{3}-r_{0}^{2}r)+\beta_{1}\beta_{2}(r^{4}+r_{0}^{4})](\alpha_{1}-\beta_{1})(\alpha_{2}-\beta_{2}),\\
&f'_{2}=\frac{1}{16}[2\alpha_{1}\alpha_{2}r^{2}+(\alpha_{1}\beta_{2}+\beta_{1}\alpha_{2})(r^{2}r_{0}+r^{3}-r_{0}^{3}+r_{0}^{2}r)+\beta_{1}\beta_{2}(r^{4}-r_{0}^{4}+2r^{2}r_{0}^{2})](\alpha_{1}+\beta_{1})(\alpha_{2}+\beta_{2}),\\
&f'_{3}=\frac{1}{16}[2\alpha_{1}\alpha_{2}r-(\alpha_{1}\beta_{2}-\beta_{1}\alpha_{2})(r^{2}+r_{0}^{2})-\beta_{1}\beta_{2}(rr_{0}^{2}+r^{3}+r_{0}^{3}-r_{0}r^{2})](\alpha_{1}+\beta_{1})(\alpha_{2}-\beta_{2}),\\
&f'_{4}=\frac{1}{16}[2\alpha_{1}\alpha_{2}r-(\alpha_{1}\beta_{2}-\beta_{1}\alpha_{2})(2rr_{0}+r^{2}-r_{0}^{2})-\beta_{1}\beta_{2}(rr_{0}^{2}+r^{3}-r_{0}^{3}+r_{0}r^{2})](\alpha_{1}-\beta_{1})(\alpha_{2}+\beta_{2}),\\
&f'_{5}=\frac{1}{16}[2\alpha_{1}\alpha_{2}r+(\alpha_{1}\beta_{2}-\beta_{1}\alpha_{2})(r^{2}+r_{0}^{2})-\beta_{1}\beta_{2}(rr_{0}^{2}+r^{3}+r_{0}^{3}-r_{0}r^{2})](\alpha_{1}-\beta_{1})(\alpha_{2}+\beta_{2}),\\
&f'_{6}=\frac{1}{16}[2\alpha_{1}\alpha_{2}r+(\alpha_{1}\beta_{2}-\beta_{1}\alpha_{2})(2rr_{0}+r^{2}-r_{0}^{2})-\beta_{1}\beta_{2}(rr_{0}^{2}+r^{3}-r_{0}^{3}+r_{0}r^{2})](\alpha_{1}+\beta_{1})(\alpha_{2}-\beta_{2}),\\
&f'_{7}=\frac{1}{16}[2\alpha_{1}\alpha_{2}-2(\alpha_{1}\beta_{2}+\beta_{1}\alpha_{2})r_{0}+2\beta_{1}\beta_{2}r_{0}^{2}](\alpha_{1}+\beta_{1})(\alpha_{2}+\beta_{2}),\\
&f'_{8}=\frac{1}{16}[2\alpha_{1}\alpha_{2}-2(\alpha_{1}\beta_{2}+\beta_{1}\alpha_{2})r+2\beta_{1}\beta_{2}r^{2}](\alpha_{1}-\beta_{1})(\alpha_{2}-\beta_{2}).
\end{split}\end{equation}

In our hyper-EPP,  our P-P-SWAP gate is used to swap the
maximally entangled states between two photon pairs, and
$\alpha_{1}=\beta_{1}=\alpha_{2}=\beta_{2}$. Thus, the fidelity and
efficiency of the P-P-SWAP gate can be expressed as
%\begin{widetext}
\begin{equation}\begin{split}
F_{SWAP}=&\frac{|f_{1}+f_{2}|^{2}}{\eta_{SWAP}},\\
\eta_{SWAP}=&\frac{1}{32}[|r^{2}+r^{2}r_{0}+r^{3}+r_{0}^{3}-r_{0}^{2}r+\frac{1}{2}(r^{4}+r_{0}^{4})|^{2}\\
&+|r^{2}+r^{2}r_{0}+r^{3}-r_{0}^{3}+r_{0}^{2}r+\frac{1}{2}(r^{4}-r_{0}^{4}+2r^{2}r_{0}^{2})|^{2}\\
&+2|r-\frac{1}{2}(rr_{0}^{2}+r^{3}+r_{0}^{3}-r_{0}r^{2})|^{2}+2|r-\frac{1}{2}(rr_{0}^{2}+r^{3}-r_{0}^{3}+r_{0}r^{2})|^{2}\\
&+|1-2r_{0}+r_{0}^{2}|^{2}+|1-2r+r^{2}|^{2}].
\end{split}\end{equation}
Here
$f_{1}=\frac{1}{16}[2r^{2}+2(r^{2}r_{0}+r^{3}-r_{0}^{3}+r_{0}^{2}r)+r^{4}-r_{0}^{4}+2r^{2}r_{0}^{2}]$
and $f_{2}=\frac{1}{8}(r_{0}-1)^{2}$. The fidelity and the
efficiency of our P-P-SWAP gate varying with the parameter
$g/\sqrt{\kappa\gamma}$ are shown in Figs.~\ref{fig11}(a) and
\ref{fig11}(b), respectively.

\end{widetext}

%\end{CJK*} %显示中文

\end{document}